\documentclass[12pt]{article}%
\usepackage[nosort]{cite}
\usepackage{graphicx}
\usepackage{multicol}
\usepackage{amsfonts}
\usepackage{amsmath}
\usepackage{amssymb}
\usepackage{amsthm}
\usepackage{heck}
\usepackage{afterpage}
\usepackage{setspace}
\usepackage{verbatim}
\usepackage{color}
\usepackage{longtable}
\usepackage{float}
\usepackage{subcaption}
\usepackage{epsfig}
\usepackage{epstopdf}
\usepackage{adjustbox}
\usepackage{tikz}
\usepackage[margin=1in]{geometry}
\usepackage{titletoc}
\usepackage{hyperref}%
\usepackage{mathrsfs}
\usepackage{dsfont}
\usepackage{algorithm}
\usepackage{algpseudocode}
\usepackage{pgfplots}
\usepackage{algorithmicx, algpseudocode,algorithm}
\usepackage{caption}
\usepackage{float}
\usepackage{makecell}
\usepackage{mathtools}
\usepackage{pdflscape}
\usepackage{array}
\usepackage{longtable}
\usepackage{comment}
\usepackage{mystuff}

\numberwithin{equation}{section}
\newsavebox{\mysavebox}

\hypersetup{colorlinks,citecolor=black,filecolor=black,linkcolor=black,urlcolor=black}

\usepackage[capitalize, noabbrev]{cleveref}
\usetikzlibrary{decorations.markings}
\usetikzlibrary{decorations.markings}
\usetikzlibrary{hobby}
\usepackage{tikz}
\usepackage{tikz-cd}
\usetikzlibrary{arrows}
\usetikzlibrary{arrows.meta}
\usetikzlibrary{arrows,decorations.pathmorphing}
\usetikzlibrary{shapes.geometric,calc,arrows, positioning,shapes.misc,decorations.markings}
\tikzset{
  big arrow/.style={
    decoration={markings,mark=at position 1 with {\arrow[scale=2,#1]{>}}},
    postaction={decorate},
    shorten >=0.4pt},
  big arrow/.default=black}

\pgfdeclarelayer{edgelayer}
\pgfdeclarelayer{nodelayer}
\pgfsetlayers{edgelayer,nodelayer,main}
\pgfplotsset{compat=1.16}
\tikzstyle{none}=[inner sep=0pt]

\renewcommand{\dd}{\mathrm{d}}
\newcommand{\ther}{\mathrm{th}}

\theoremstyle{definition}

\crefname{thm}{Theorem}{Theorems}
\crefname{prop}{Proposition}{Propositions}
\crefname{defn}{Definition}{Definitions}
\crefname{lem}{Lemma}{Lemmas}

\tikzstyle{NodeCross}=[draw, shape=circle, cross out, inner sep=0pt, minimum size=6pt,line width=0.25mm]
\tikzstyle{Circle}=[draw, shape=circle, black, fill=black, inner sep=0pt, minimum size=6pt]
\tikzstyle{circle}=[draw, shape=circle, black, fill=black, inner sep=0pt, minimum size=16pt]
\tikzstyle{Star}=[draw, shape=star, fill=black, star points=8, inner sep=0pt, minimum size=8pt]
\tikzstyle{CircleRed}=[draw, shape=circle, black, fill=red, inner sep=0pt, minimum size=6pt]
\tikzstyle{StarP}=[draw={rgb,255: red,128; green,0; blue,128}, shape=star, fill={rgb,256: red,128; green,0; blue,128}, star points=8, inner sep=0pt, minimum size=12pt]
\tikzstyle{ShadedCircRed}=[draw=red, shape=circle, fill={rgb, 255: red,255; green,114; blue, 118}, inner sep=0pt, minimum size=80pt, line width=0.5mm, fill opacity=0.2]
\tikzstyle{ShadedCircRed2}=[draw=red, shape=circle, fill={rgb, 255: red,255; green,114; blue, 118}, inner sep=0pt, minimum size=10pt]
\tikzstyle{ShadedCircRed3}=[draw=black, shape=rectangle, fill={rgb, 255: red,255; green,114; blue, 118}, inner sep=0pt, minimum size=113pt, line width=0.25mm]
\tikzstyle{ShadedCirc}=[draw=red, shape=circle, fill=white, inner sep=0pt, minimum size=45pt,  fill opacity=1.0,  line width=0.5mm]
\tikzstyle{CircleBlue}=[draw, shape=circle, fill=blue, inner sep=0pt, minimum size=6pt]
\tikzstyle{BigCirclePurple}=[draw, shape=circle, fill={rgb,255: red,191; green,0; blue,191}, inner sep=0pt, minimum size=12pt]
\tikzstyle{CirclePurple}=[draw, shape=circle, fill={rgb,255: red,191; green,0; blue,191}, inner sep=0pt, minimum size=5pt]
\tikzstyle{EmptyCircle}=[draw, shape=circle, inner sep=0pt, minimum size=4pt]
\tikzstyle{GreenCircle}=[draw, shape=circle,  fill={rgb,255: red,80; green,200; blue,120}, inner sep=0pt, minimum size=8pt]
\tikzstyle{BrownCircle}=[draw, shape=circle,  fill={rgb,255: red,210; green,105; blue,30}, inner sep=0pt, minimum size=8pt]
\tikzstyle{CirclePurpleSmall}=[draw, shape=circle, fill={rgb,255: red,191; green,0; blue,191}, inner sep=0pt, minimum size=4pt]
\tikzstyle{BigCircleGreen}=[draw, shape=circle, fill={rgb,255: red,0; green,191; blue,0}, inner sep=0pt, minimum size=12pt]
\tikzstyle{BigCircleBlue}=[draw, shape=circle, fill={rgb,255: red,0; green,0; blue,191}, inner sep=0pt, minimum size=12pt]
\tikzstyle{BigCircleRed}=[draw, shape=circle, fill={rgb,255: red,191; green,0; blue,0}, inner sep=0pt, minimum size=12pt]
\tikzstyle{BrownCircleSmall}=[draw, shape=circle,  fill={rgb,255: red,210; green,105; blue,30}, inner sep=0pt, minimum size=6pt]
\tikzstyle{SmallCircleBrown}=[draw, shape=circle,  fill={rgb,255: red,210; green,105; blue,30}, inner sep=0pt, minimum size=5pt]
\tikzstyle{SmallCircleRed}=[draw, shape=circle, fill={rgb,255: red,191; green,0; blue,0}, inner sep=0pt, minimum size=6pt]
\tikzstyle{DashedLine}=[-, densely dashed, line width=0.25mm]
\tikzstyle{DottedLine}=[-, dotted, line width=0.25mm]
\tikzstyle{ThickLine}=[-, line width=0.25mm]
\tikzstyle{ArrowLineRight}=[-, -{Stealth[scale=1.25]}, line width=0.25mm, scale=5]
\tikzstyle{ArrowLineRed}=[-, draw={rgb,255: red,191; green,0; blue,0}, -{Stealth[scale=1.75]}, line width=0.1mm, scale=5]
\tikzstyle{RedLine}=[-, draw={rgb,255: red,191; green,0; blue,0}, fill=none, line width=0.5mm]
\tikzstyle{DashedLineThin}=[-, densely dashed, line width=0.125mm, fill=none, draw=black]
\tikzstyle{DottedRed}=[-, dotted, draw={rgb,255: red,191; green,0; blue,0}, fill=none, line width=0.25mm]
\tikzstyle{DashedRed}=[-, densely dashed, draw={rgb,255: red,191; green,0; blue,0}, fill=none, line width=0.25mm]
\tikzstyle{BlueLine}=[-, draw={rgb,255: red,0; green,0; blue,191}, fill=none, line width=0.5mm]
\tikzstyle{ArrowLineBlue}=[-, draw={rgb,255: red,0; green,0; blue,191}, -{Stealth[scale=1.75]}, line width=0.1mm, scale=5]
\tikzstyle{GreenDoubleArrow}=[<->, draw={rgb,155: red,0; green,255; blue,0},  line width= 0.5mm, scale=5]
\tikzstyle{RedDoubleArrow}=[<->, draw={rgb,255: red,255; green,0; blue,0},  line width= 0.5mm, scale=5]
\tikzstyle{BlueDottedLight}=[-, dotted, draw={rgb,255: red,0; green,0; blue,191}, fill=none, line width=0.3mm]
\tikzstyle{BrownLine}=[-, draw={rgb,255: red,210; green,105; blue,30}, fill=none, line width=0.5mm]
\tikzstyle{DottedRed}=[-, dotted, draw={rgb,255: red,191; green,0; blue,0}, fill=none, dotted, line width=0.5mm]
\tikzstyle{DottedPurple}=[-, dotted, draw={rgb,255: red,191; green,0; blue,191}, fill=none, dotted, line width=0.5mm]
\tikzstyle{BlueDottedLight}=[-, dotted, draw={rgb,255: red,0; green,0; blue,191}, fill=none, line width=0.5mm]
\tikzstyle{ArrowLinePurple}=[-, draw={rgb,255: red,191; green,0; blue,191}, -{Stealth[scale=1.75]}, line width=0.5mm, scale=5]
\tikzstyle{DashedLineGreen}=[-, densely dashed, draw={rgb,255: red,74; green,103; blue,65}, line width=0.25mm]
\tikzstyle{LineGreen}=[-, draw={rgb,255: red, 74; green,200; blue,65}, line width=0.5mm]
\tikzstyle{ArrowLineGreen}=[-, draw={rgb,255: red,0; green,191; blue,0}, -{Stealth[scale=1.75]}, line width=0.5mm, scale=5]
\tikzstyle{GreenLine}=[-, draw={rgb,255: red,0; green,191; blue,0}, fill=none, line width=0.5mm]
\tikzstyle{PurpleLine}=[-, draw={rgb,255: red,191; green,0; blue,191}, fill=none, line width=0.5mm]
\tikzstyle{PPurpleLine}=[-, draw={rgb,255: red,191; green,0; blue,191}, fill=none, line width=2.5mm]
\tikzstyle{DPurpleLine}=[-, dotted, draw={rgb,255: red,191; green,0; blue,191}, fill=none, line width=0.5mm]
\tikzstyle{SBrownLine}=[-, draw={rgb,255: red,191; green,0; blue,191}, fill=none, opacity=0.35, line width=2.5mm]
\tikzstyle{DottedBlue}=[-, dotted, draw=blue, fill=none, dotted, line width=0.5mm]
\tikzstyle{DashedPurpleLine}=[-, densely dashed, draw={rgb,255: red,191; green,0; blue,191}, fill=none, line width=0.5mm]
\tikzstyle{SmallCircleBlue}=[draw, shape=circle, fill=blue, inner sep=0pt, minimum size=5pt]
\tikzstyle{SmallCirclePurple}=[draw, shape=circle, fill={rgb,255: red,191; green,0; blue,191}, inner sep=0pt, minimum size=5pt]
\tikzset{snake it/.style={decorate, decoration=snake}}

\tikzset{
dashstar/.style={
 dash pattern=on 5pt off 5pt,
 postaction={
  decorate,
  decoration={
   markings,
   mark=between positions 9pt and 1 step 10pt with {
     \node[color=red] {*};
   }
  }
 }
},
dashstarstar/.style={ 
 dash pattern=on 5pt off 10pt,
 postaction={
   decorate,
   decoration={
     markings,
     mark=between positions 10pt and 1
          step 15pt
           with {
            \node at (-2pt,0pt) {\pgfuseplotmark{star}};
            \node at (2pt,0pt) {\pgfuseplotmark{star}};
           }
   }
 }
}
}

\begin{document}

\date{April 2026}

\title{Generalized Complexity Distances \\[4mm] and Non-Invertible Symmetries}

\institution{PENN}{\centerline{$^{1}$Department of Physics and Astronomy, University of Pennsylvania, Philadelphia, PA 19104, USA}}
\institution{PENNmath}{\centerline{$^{2}$Department of Mathematics, University of Pennsylvania, Philadelphia, PA 19104, USA}}

\authors{
Jonathan J. Heckman\worksat{\PENN, \PENNmath}\footnote{e-mail: \texttt{jheckman@sas.upenn.edu}},
Rebecca J. Hicks\worksat{\PENN}\footnote{e-mail: \texttt{rjhicks@sas.upenn.edu}}, and
Chitraang Murdia\worksat{\PENN}\footnote{e-mail: \texttt{murdia@sas.upenn.edu}}
}

\abstract{
Non-invertible symmetries of a quantum field theory (QFT) are a natural generalization of unitary symmetries, but in which the product of operators does not satisfy a group multiplication law.
We show that such symmetry operations on states define a collection of quantum gates for a parallel quantum computation scheme that includes post-selection / projection as a gate.
Structures such as gate complexity and more geometric complexity measures generalize to this setting.
We provide a class of distance / distinguishability measures that extend the standard notion of distance for Lie groups to both continuous and discrete non-invertible symmetries, as well as more general linear combinations of unitary quantum gates. We illustrate these considerations by
computing the distance between non-invertible symmetries in some 4D and 2D QFTs.
We find that the simple objects of a symmetry category can be highly complex computationally.}

\maketitle

\enlargethispage{\baselineskip}

\setcounter{tocdepth}{2}

\tableofcontents

\newpage

\section{Introduction}

Symmetry principles provide important constraints in a broad range of classical and quantum systems.
In the quantum setting, symmetry is commonly phrased in terms of a suitable unitary representation of a
group \cite{Wigner:1931}. This representation acts on the states of the quantum system, leading to non-trivial selection rules.

Recent developments in quantum field theory have established that there are deep topological structures connected
with symmetries \cite{Gaiotto:2014kfa}. Perhaps surprisingly, such symmetries
can relax the standard group multiplication law. An example of this sort are
non-invertible symmetries, with fusion rule:
\begin{equation}\label{eq:NONINV}
\mathcal{X}_i \mathcal{X}_j = \underset{k}{\sum} N_{ij}^{k} \mathcal{X}_k.
\end{equation}
Here, the $\mathcal{X}_a$ denote non-invertible symmetry operators; they can be viewed as the path integral of a topological quantum field theory.
The fusion coefficients $N_{ij}^{k}$ can also be interpreted as the path integrals of quantum field
theories,\footnote{Though of course they can also be just numbers.} leading to a rich generalization of ``conventional symmetries.''
There is by now an extensive literature on how non-invertible symmetries appear in many quantum systems,
see e.g., \cite{Shao:2023gho} for a review.

One of the curious features of non-invertible symmetries is that they are not specified by a unitary representation of a group, i.e., they are not implemented by unitary operators. As such, they alter the norm of quantum states, making some of their information theoretic properties more subtle to interpret.\footnote{See \cite{Okada:2024qmk, Benini:2025lav, Ortiz:2025psr, Demulder:2026vbn, Demulder:2026bey, Bartsch:2026wqq} for interpretations of non-invertible symmetries as isometric quantum channels.}

One of the aims of the present work will be to interpret non-invertible symmetries in the broader setting of quantum computation,\footnote{See \cite{Nielsen:2012yss} for a review of quantum
computation.}  where we view them as a special case of a more general class of
quantum gates known as linear combinations of unitary operators (LCUs).
Quantum computations are typically formulated by specifying a
basis of qubit / qudit states $\ket{j_1} \otimes \ket{j_2} \otimes \dots \ket{j_N} \in \mathcal{H}_{\mathrm{qubits}}$ which are acted upon by a collection of prescribed unitary operators drawn from a gate set:
\begin{equation}
G = \{g_1,...,g_N \}.
\end{equation}
In a quantum computation, one envisions a choice of time-dependent Hamiltonian built up from an appropriate sequence of $g_i$'s being applied. The complexity class associated with bounded-error quantum polynomial time computations is referred to as $\mathbf{BQP}$.

\begin{figure}[t!]
    \centering
    \includegraphics[width=0.75\textwidth]{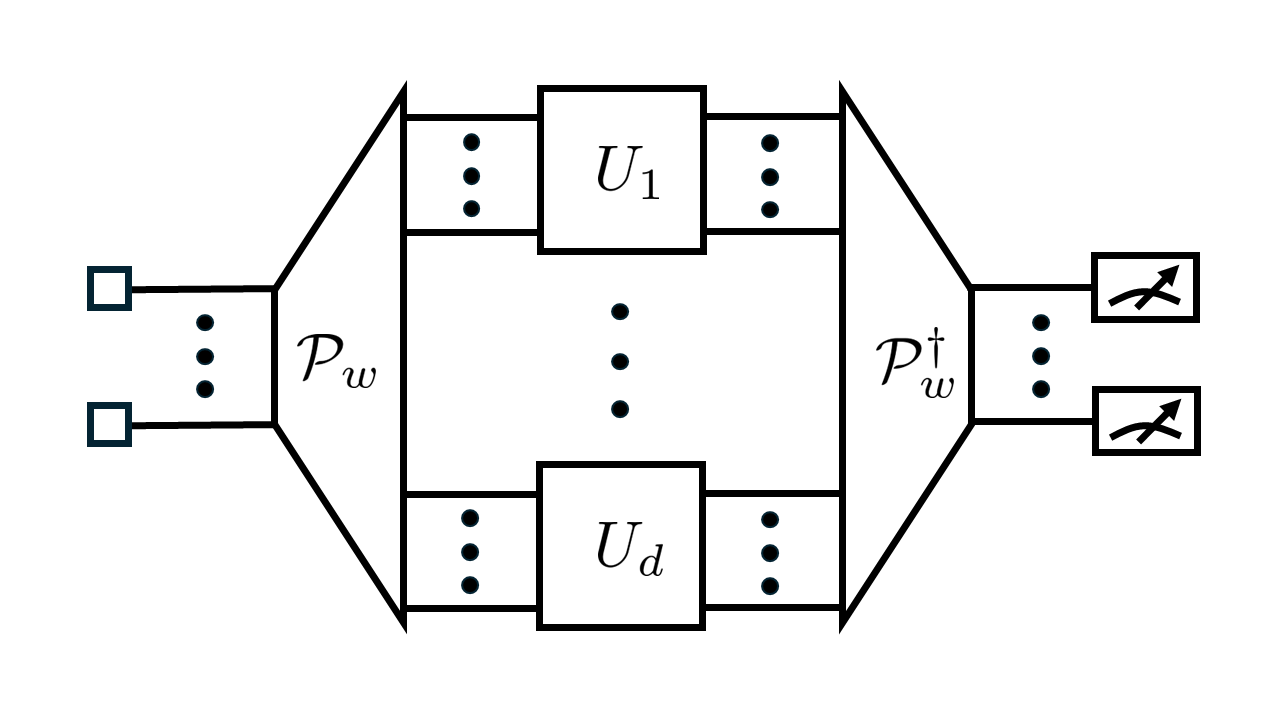}
    \caption{Depiction of a set of parallel quantum computations by unitary operators $U_1,...,U_d$ padded by ancillary qubits / qudits. The linear combination $w_1 U_1 + ... + w_d U_d$ is obtained by sandwiching these parallel computations inside of the prepare operator $\mathcal{P}_w$ and its hermitian conjugate $\mathcal{P}^{\dag}_{w}$ which acts on these ancillary qubits / qudits. At the right, the outcome of this computation is read out on a measurement device.}
    \label{fig:ParallelComp}
\end{figure}

One can also entertain a far broader class of operators beyond just unitary ones,
and it has been appreciated for some time that this provides a more flexible
notion of quantum gates and quantum computations.
Indeed, allowing for post-selection, one can also consider linear combinations of unitary gates such as:
\begin{equation}
O_{w} \equiv  \underset{i}{\sum} w_i g_i,\,\,\,\text{with}\,\,\, \underset{i}{\sum} w_i = 1.
\end{equation}
for $w_i \in [0,1]$ normalized weights. This broader class of operators furnishes us with
linear combinations of unitary operators (LCUs). Observe that in general,
the product of such $O_{w}$ will obey a fusion rule,
more akin to line (\ref{eq:NONINV}) than a group law. A natural way to visualize such operations
is to consider a collection of parallel quantum computations which is then projected to a single computer. Including projection / post-selection
extends $\mathbf{BQP}$ to $\mathbf{PostBQP} \supset \mathbf{BQP}$ \cite{Aaronson:2004voc}.
Thus, implementing a given LCU certainly lies in the complexity class $\mathbf{PostBQP}$,  however it might not be $\mathbf{PostBQP}$-complete.
In this broader landscape of quantum operations, non-invertible symmetries
simply amount to a special class of LCUs; one can entertain them as enlarging the gate set of $\mathbf{BQP}$ rather naturally. See figure \ref{fig:ParallelComp} for a depiction of this parallelization and post-selection.

From this perspective, it is natural to ask about efficient implementations of gate sets and operations on quantum states,
i.e., the quantum complexity of an operation. One approach to this question is to simply attempt to find a ``minimal'' sequence
of gates which implements a given operation, though it is notoriously challenging to find optimal gate sets. In the context of
operations constructed via unitary operators, there is an elegant geometric formulation of complexity in terms of Nielsen's distance
measure on the Lie group $U(2^N)$.\footnote{See in particular \cite{Nielsen:2005mkt, Nielsen:2006cea, Nielsen:2006mgv, Dowling:2006tnk, Gu:2008rsv}, and the overviews provided in \cite{Brown:2019rox, Acevedo:2025juf}}

We propose to extend this geometric formulation of complexity to cover the broader case of LCUs. In particular, this allows us to define
a natural notion of a distance / proximity between non-invertible symmetries in any quantum field theory. The main idea underlying
our approach is to fix a reference mixed state $\rho$ and its canonical purification $\vert \psi_{\rho} \rangle$. The action
of an LCU on $\rho$ naturally extends to the purification, and so we study the quantum distinguishability of states acted on by an LCU. This distinguishability provides a clear information theoretic notion of ``proximity'' which applies equally well to both
continuous and discrete symmetries, be they invertible or non-invertible. Our preliminary information-theoretic motivated analysis returns
a metric which generalizes the Killing metric of a Lie group / symmetric space. From this, we can further extrapolate to produce
complexity metrics. As such, it provides the desired generalization of Nielsen's
complexity measure in this broader setting.

We illustrate these general considerations with a few representative examples, including 4D $O(2)$ gauge theory (i.e., Maxwell theory with a gauged charge conjugation symmetry), 2D rational conformal field theories (RCFTs), and the $N$-fold symmetric product of a 2D CFT. A general feature of our analysis for non-invertible symmetries is that in certain limits, they tend to have \textit{maximal} gate complexity.
As such, the ``simple objects'' of the symmetry category turn out to be maximally complex computationally!

The rest of this paper is organized as follows. In section \ref{sec:GATES} we interpret non-invertible symmetries as LCUs and, in particular, show how they specify a natural class of parallelized quantum computations with post-selection. In section \ref{sec:DISTANCE} we define a notion of distance for symmetry operators and show that it generalizes Nielsen's
complexity measure. We turn to explicit examples in section \ref{sec:EXAMPLES}, and in section \ref{sec:CONC} we present our conclusions. Some additional review material is deferred to the Appendices.

\section{Quantum Gates and Non-Invertible Symmetries} \label{sec:GATES}

In this section we interpret non-invertible symmetries as gates in a quantum computation with post-selection.
With this established, we turn to a discussion of computation with linear combinations of unitary gates, and discuss some
basic properties of gate complexity for linear combinations of unitary operators (LCUs).

To set terminology, we refer to a symmetry operator as any topological operator of a quantum field theory (QFT) which commutes with the stress energy tensor. As a topological operator, any candidate symmetry operator can implicitly be defined by a topological quantum field theory with a path integral over its degrees of freedom. We denote this schematically as:
\begin{equation}\label{eq:Ndef}
\mathcal{X}[\Phi] \sim \int [da] \exp \left(i S_{\mathrm{TFT}}[a,\Phi] \right),
\end{equation}
where $\Phi$ denotes possible background fields of the ambient QFT, and the TFT is (implicitly) supported on a manifold $\Sigma$ contained within the spacetime supporting the ambient QFT. The special case of a non-invertible symmetry is one for which its fusion rules with some other operators is non-trivial:
\begin{equation}
\mathcal{X}_{i} \mathcal{X}_{j} = \underset{k}{\sum} N^{k}_{ij} \mathcal{X}_k,
\end{equation}
where as mentioned in the introduction, the $N^{k}_{ij}$ can themselves be
non-trivial path integrals over (possibly lower-dimensional) TFTs. Especially in the case of $D > 2$ QFTs, much of the non-trivial structure
present in fusion rules is concentrated on lower-dimensional strata, i.e., the fusion rules
primarily involve ``condensation defects'' (see e.g., \cite{Gaiotto:2019xmp, Roumpedakis:2022aik}).

For our purposes, the appearance of a suitable path integral phase means
intuitively that the operators of line (\ref{eq:Ndef}) can always be interpreted as a 
sum over unitary operators with real, positive weights which sum to one.
This is actually to be expected since for \textit{any} bounded operator acting on a $C^{\ast}$ algebra,
there is a corresponding decomposition into a weighted sum of unitary operators belonging to the \textit{same} $C^{\ast}$ algebra.
Indeed, from \cite{Pedersen1985}, we have that for an operator $A$
with $\vert \vert A \vert \vert \leq 1 - 2d^{-1}$, and $d \in \mathbb{N}$,
then there exist unitary operators $U_1,...,U_d$ such that:
\begin{equation}\label{eq:AVGSUM}
A = \frac{1}{d} \left(U_1 + ... + U_d \right).
\end{equation}
Observe that for any bounded operator, we can always rescale the norm, so this is a rather general
result.\footnote{Let us present an illustrative example of a weighted average. 
Consider a bounded operator $A$ of a $C^{\ast}$ algebra. Observe that we can present $A$ as:
\begin{equation}
A = B + \widetilde{B},
\end{equation}
with $B^\ast = B$ and $\widetilde{B}^{\ast} = - \widetilde{B}$ (for a Hilbert space the $\ast$ operation is simply
the usual hermitian conjugation operation $\dag$). Let us assume that $A$ has eigenvalues 
with norm smaller than one (or alternatively consider a rescaled version of $A$). Note that the eigenvalues of $B$ and $\widetilde{B}$ 
also have norm less than one. Consequently, we can decompose $B$ and $\widetilde{B}$ as:
\begin{align}
B & = \frac{1}{2} \left(B + i \sqrt{\mathbb{I} - B^{\ast} B} \right) + \frac{1}{2} \left(B - i \sqrt{\mathbb{I} - B^{\ast} B} \right) \\
\widetilde{B} & = \frac{1}{2} \left(\widetilde{B} +  \sqrt{\mathbb{I} + \widetilde{B}^{\ast} \widetilde{B}} \right) + \frac{1}{2} \left(\widetilde{B} -  \sqrt{\mathbb{I} + \widetilde{B}^{\ast} \widetilde{B}} \right),
\end{align}
where each term in parentheses manifestly satisfies $U^{\ast} U = U U^{\ast} = \mathbb{I}$, in the obvious notation.
By inspection, we see that for this $A$, we can write it as:
\begin{equation}
A = \frac{1}{2} \left( U_1 + U_2 + U_3 + U_4 \right),
\end{equation}
where each $U_j$ is unitary, and the weights sum to $2$. 
Consequently, the rescaled operator $A^{\prime} = A/2$ can be written as a normalized weighted sum.
Observe that the norm of this operator is $\vert \vert A^{\prime} \vert \vert  = \vert \vert A \vert \vert / 2 < 1/2$.
For operators with norm closer to one, averaging over more unitaries is required.
\label{foot:DFWagain}
}

As an additional comment, note that in the specific context of a symmetry operator of a quantum field theory,
i.e., one that commutes with the stress energy tensor, there is a priori no reason for the $U_j$'s to also be symmetry operators, but there are many situations where one can trade the ``non-invertible'' basis of symmetry
operators for invertible ones. One might then ask if we can equally well work with a collection of unitary operators
which commute with the Hamiltonian, why bother with non-invertible symmetries at all?
In general, these unitary operators are not the simple objects of the symmetry category,
that role being reserved for the non-invertible symmetry operators.\footnote{
More precisely, instead of viewing an operator as acting at a fixed time, we can instead consider a
timelike insertion of the operator, thus constructing a defect Hilbert space. In the case where
the unitary operator $U$ is not a simple object of the symmetry category,
the defect Hilbert space $\mathcal{H}_U$ obtained from inserting
this operator will not have a unique identity operator. In this sense, the
non-invertible symmetries are still the more primitive / fundamental objects. See the review \cite{Shao:2023gho} for a helpful discussion on this point. We thank X. Yu for correspondence.\label{foot:DFWstyle}}

In a general QFT there can be a rather rich categorical structure of symmetries which relaxes many
of the standard group-like multiplication rules. Non-trivial fusion rules, as captured by
non-invertible symmetries is but one way in which these structures can get relaxed. For example,
pulling operators through one another in the spacetime leads to non-trivial attachments by disorder operators, i.e., non-trivial associators. These are structures directly tied to the spacetime, and so there is of course still a formal notion of each presentation of the operators acting on a Hilbert space of states.\footnote{The representation theory can be more intricate. For example, in the case of 2D CFTs, Ocneanu's tube algebra is the relevant generalization. See also \cite{Cordova:2024iti} for a recent discussion for massive 2D QFTs.}

To give an example, consider the Fibonacci
anyon system with elements $\mathbb{I},\tau$ and fusion rule $\tau \times \tau = \mathbb{I} + \tau$,
where we assume that $\tau$ acts on a suitable anyon Hilbert space with $\tau = \tau^{\dag}$.\footnote{This operator
can be realized in terms of an explicit lattice system \cite{Feiguin:2006ydp}.}
Observe that we can build a unitary operator $U_{\tau}$ as:
\begin{equation}
U_{\tau} = -\frac{i}{ \sqrt{2}} \left(\mathbb{I} + b \tau \right),
\end{equation}
where the condition that $U_{\tau}$ is unitary requires $\vert b \vert^2 + b + b^{\ast} = 0$, i.e., $b = e^{2 \pi i / 3}$.
Then, we can express $\tau$ as:
\begin{equation}
\tau = \sqrt{\frac{2}{3}} \left(U_{\tau} + U_{\tau}^{\dag} \right).
\end{equation}
The normalization factor here is $\sqrt{2/3}$ (rather than $1/2$ as in footnote \ref{foot:DFWagain})
because the eigenvalues of $\tau$ are $(1 \pm \sqrt{5}) / 2$.

Observe that since $\mathbb{I}$ and $\tau$ both commute
with the Hamiltonian, $U_{\tau}$ is also a symmetry operator.
That being said, it is not a simple object of the symmetry category since the defect
Hilbert space (obtained by inserting a timelike $U_{\tau}$ line) does not have a unique identity operator, whereas the simple objects will.\footnote{See footnote \ref{foot:DFWstyle}.}
Nevertheless, nothing prevents us from introducing such a formal decomposition.

More generally, in situations where the fusion coefficients also specify non-trivial TFTs,
this just specifies a further product of operators.\footnote{In the context of generalized symmetries where there is a stratification according to the dimension of support of a topological operator, it is
natural to view the zero-form symmetries (codimension one) as specifying the primary gate operation
and the $p$-form symmetries for $p > 0$ as further controls on a computation. We defer a full treatment
of this issue to future work.} As such, we have at a rather general level the statement that we can take a given bounded operator,
rescale it to have norm less than one. Then, it suffices to consider operators given by
linear combinations with weights $w_{i} \in [0,1]$:
\begin{equation}\label{eq:OWOW}
O_{w} = w_1 U_1 + ... + w_d U_d \,\,\, \text{with} \,\,\, \underset{i}{\sum}{w_i} = 1.
\end{equation}
The class of operators specified in this way are known as LCUs, where we have restricted to the computationally natural choice of all weights non-negative and suitably normalized. This choice for the $w_i$'s is amenable to a computational interpretation. Indeed, here we are treating operators that differ by an overall multiplicate constant to be equivalent, since their action on any state only differs in the overall normalization.

Our aim will be to give a natural interpretation of such LCUs, eventually
applied to the case of non-invertible symmetries in QFTs. Now, in the case of quantum field theories, the underlying Hilbert space is infinite-dimensional. At a practical level, however, we can opt to truncate the distinguishability of states by some tolerance $\varepsilon > 0$, or even more crudely, simply assume a suitable finite-dimensional regularization (e.g., a lattice regularization).
With these caveats spelled out, our aim will be to understand non-invertible symmetries as a special class of LCUs which enact quantum gates.

We can visualize the LCUs as specifying a collection of parallel
operations in a bigger Hilbert space which we then project down to the original set. Along these lines,
suppose we have a linear combination such as $O_{w}$ in line (\ref{eq:OWOW}). To deal with the $U_i$'s appearing in this sum, we introduce an ancillary qudit Hilbert space $\mathcal{H}_{\mathrm{qudit}}$ spanned by states $\vert k \rangle$
for $k = 0,...,d-1$. We can then consider constructing a unitary operator on $\mathcal{H} \otimes \mathcal{H}_{\mathrm{qudit}}$ given by:
\begin{equation}
\mathbb{U} \equiv \mathrm{diag}(U_1 ,..., U_d) = \sum_{k=0}^{d-1} U_{k+1} \otimes \ket{k} \bra{k} \, ,
\end{equation}
which by inspection amounts to $d$ parallel operations. To account for the different weights appearing in the linear combination $O_{w}$, introduce a unitary ``prepare operator''
\begin{equation}
\mathcal{P}_{w}: \mathcal{H}_{\mathrm{qudit}} \rightarrow \mathcal{H}_{\mathrm{qudit}}
\end{equation}
which acts on $\vert 0 \rangle \in \mathcal{H}_{\mathrm{qudit}}$ as:
\begin{equation}
\mathcal{P}_{w} \vert 0 \rangle = \sum_{k=0}^{d-1} \sqrt{w_{k+1}} \ket{k} \, .
\end{equation}
where for our purposes, its action on the other qudit states is unimportant since the only requirement is that $\mathcal{P}_w$ is unitary. $\mathcal{P}_{w}$ extends to $\mathcal{H} \otimes \mathcal{H}_{\mathrm{qudit}}$ by tensoring with the identity operator on $\mathcal{H}$.
Observe that we can post-select the ancilla qudit to be in the $\ket{0}$ state to obtain the desired final state:
\begin{equation}\label{eq:READOFF}
\begin{split}
    \ket{\psi_F} &= \bra{0} \mathcal{P}_w^\dagger \left( \sum_{k=0}^{d-1} U_{k+1} \otimes \ket{k} \bra{k} \right) \mathcal{P}_w \ket{\psi_I} \otimes \ket{0} \\
    &= \sum_{j=1}^{d} w_{j} U_{j} \ket{\psi_I} \, ,
\end{split}
\end{equation}
i.e., this implements the action of $O_w$ on a state of the original Hilbert space $\mathcal{H}$.
We visualize this as a collection of parallel quantum computations which are then recombined via post-selection.
Multiple operations with LCUs work similarly; we can build a general composition by introducing a collection of weights and choices of
$\mathbb{U}$'s. Denoting this sequence of weights as $w^{(m)}_{i}$, we construct a sequence of prepare operators $\mathcal{P}_{(m)}$.
Likewise, denote the corresponding parallelized unitaries as $\mathbb{U}_{(m)}$. Then, we produce the desired computation from the composition:
\begin{equation}
(\mathcal{P}_{(M)}^{\dag} \mathbb{U}_{(M)} \mathcal{P}_{(M)} ) \circ ... \circ (\mathcal{P}_{(1)}^{\dag} \mathbb{U}_{(1)} \mathcal{P}_{(1)} ),
\end{equation}
for $M$ such operations.

To better understand the quantum computational aspects of our setup, we now
review some general quantum computational features of LCUs.
From this perspective, non-invertible symmetries comprise a particular well-motivated special case of LCUs which can be viewed as supplementing a
standard gate set for quantum computation.

\subsection{Gate Complexity}

We now discuss the gate complexity of unitary and LCU gates. We begin with a set of unitary quantum gates $G = \{ g_1, g_2, \dots g_N \}$. These gates are the fundamental building block of any quantum computation protocol. We assume that $G^\dagger = G$ i.e., if $g \in G$ then $g^\dagger \in G$ as well.
The gate complexity of a unitary operator $U$ is defined as the minimal number of gates needed to prepare it. In other words, it is the smallest $n$ for which
\begin{equation}
    U = g_{i_1} g_{i_2} \dots g_{i_n} + O(\varepsilon) \, .
\end{equation}
where $\varepsilon$ characterizes the error we are willing to tolerate in our preparation.
We denote this as
\begin{equation}
    \mathcal{C}(U) = n \, .
\end{equation}

We want to extend this definition to non-unitary operators. We begin by considering the operator
\begin{equation}
    O_2 = \frac{1}{2} \left( g_1 + g_2 \right) \, .
\end{equation}
Clearly, this operator is not unitary. Unlike the unitary operator $U$, this operator cannot be constructed by a sequential action of unitary gates. However, if we allow for post-selection on ancilla qubits, this operator can be constructed as follows.
Consider the initial state $\ket{\psi_I}$. To this state, we append a qubit in state $\ket{0}$ and then act with the unitary operator
\begin{equation}
\label{eq:g1g2_select}
    g_1 \otimes \ket{+} \bra{+} + g_2 \otimes \ket{-} \bra{-} \, ,
\end{equation}
where $\ket{+} = \frac{1}{\sqrt{2}}(\ket{0}+\ket{1})$ and $\ket{-} = \frac{1}{\sqrt{2}}(\ket{0}-\ket{1})$ are the $\sigma_x$ eigenstates\footnote{These states can be easily prepared using the Hadamard and NOT gates, so state preparation does not contribute significantly to the gate complexity.}. Finally, we post-select for the ancilla to be in state $\ket{0}$ to obtain the desired final state
\begin{equation}
    \ket{\psi_F} = \bra{0} \big( g_1 \otimes \ket{+} \bra{+} + g_2 \otimes \ket{-} \bra{-} \big) \ket{\psi_I} \otimes \ket{0} = O_2 \ket{\psi_I} \, .
\end{equation}
Since this post-selection fails with probability $\frac{1}{2}$, this operation needs to be performed twice on average to get the desired output.\footnote{Here, we are assuming that $g_1,g_2$ have sufficiently random matrix elements so that $\abs{\bra{\psi_I} g_1^\dagger g_2 \ket{\psi_I}} \ll 1$. If this is not the case, the probability is $\frac{1}{4} \bra{\psi_I} (g_1+g_2)^\dagger (g_1 + g_2) \ket{\psi_I} = \frac{1}{2} \left(1 + \text{Re}\left(\bra{\psi_I} g_1^\dagger g_2 \ket{\psi_I} \right) \right)$.\label{foot:DFWDFW}}

Note that the procedure we have described above is a particularly simple instance of an algorithm for implementing an LCU \cite{Childs:2012gwh, Low:2016znh}.
The unitary operator in \eqref{eq:g1g2_select} is known as the select operator.
The standard way of implementing this operator is to apply a sequence of controlled-$g$ operators
\begin{equation}
\begin{split}
    \big( g_1 \otimes \ket{+} \bra{+} + g_2 \otimes \ket{-} \bra{-} \big) = \big( \mathbb{I} &\otimes \ket{+} \bra{+} + g_2 \otimes \ket{-} \bra{-} \big) \\
    &\times\big( g_1 \otimes \ket{+} \bra{+} + \mathbb{I} \otimes \ket{-} \bra{-} \big) \, .
\end{split}
\end{equation}
We will assume that the gate complexity of a single controlled-$g$ operator is $\mathcal{C}_g$. The specific value of this quantity depends of the details of the theory and the gate set $G$ \cite{Gosset:2013ojr,Babbush:2018ywg}.

In the above we have used a post-selection protocol that fails with some probability $p$, and we have focused on the special case $p = 1/2$.
This means that the procedure must be performed $1/p$ times on average to get the desired result.
If we allow for an error tolerance, there are deterministic algorithms known as quantum amplitude amplification that can replace this post-selection step \cite{Brassard:2000xvp}. These algorithms are a modification of Grover's search algorithm and only need $1/\sqrt{p}$ iterations of the LCU subroutine.
Thus, the complexity of the operator $O_2$ is
\begin{equation}
    \mathcal{C} \left( O_2 \right) \leq \sqrt{2} \times 2 \mathcal{C}_g \, ,
\end{equation}
where the first factor of $\sqrt{2}$ comes from amplitude amplification and the second factor of $2 \mathcal{C}_g$ comes from the LCU subroutine.
Note that this prescription only provides an upper bound on the gate complexity because there could be a simpler procedure to implement $O_2$. However, for a large class of $O_2$'s we expect that this upper bound matches the complexity.

Consider next an arbitrary Hermitian operator $O$, where we assume $\vert \vert O \vert \vert < 1$,
which can always be achieved by rescaling of $O$. In this case, $O$ can be decomposed
into a linear combination of $U_{O}$ and its Hermitian conjugate:\footnote{See footnote \ref{foot:DFWagain}.}
\begin{equation}
    O = \frac{1}{2} \left( U_{O} + U_{O}^\dagger \right) \, .
\end{equation}
The unitary $U_{O}$ can be implemented using $\mathcal{C} \qty(U_{O})$ gates.
Thus, we can use above procedure to prepare $O$ by using the unitary gates along with
post-selection / amplitude amplification on a single qubit. This provides an upper bound on the complexity of $O$
\begin{equation}
    \mathcal{C}(O) \leq 2 \sqrt{2} \mathcal{C} \qty(U_{O}) \mathcal{C}_g \, .
\end{equation}

Let us further generalize this procedure to operators of the form
\begin{equation}
    O_{d} = \frac{1}{d} \sum_{k=1}^{d} g_{i_k} \, .
\end{equation}
In this case, we append the initial state with an ancilla qudit in state $\ket{0}$ and then act with the select operator
\begin{equation}
    \sum_{k=0}^{d-1} g_{i_{k+1}} \otimes \ket{k_f} \bra{k_f} \, ,
\end{equation}
where $\ket{k_f} = \frac{1}{\sqrt{d}} \sum_{j=0}^{d-1} e^{2 \pi i j k/d} \ket{j}$ with $k \in \{0, 1, \dots d-1 \}$ are the Fourier basis eigenstates. Finally, we post-select the ancilla to be in state $\ket{0}$ to obtain the desired final state $\ket{\psi_F} = O_d \ket{\psi_I}$.
It follows that
\begin{equation}
    \mathcal{C}(O_d) \leq d^{3/2} \mathcal{C}_g \, .
\end{equation}

So, for bounded operators of the form appearing in line (\ref{eq:AVGSUM}), i.e., those which can be written as a linear combination of unitaries of the form:
\begin{equation}
    O_{d} = \frac{1}{d} \sum_{j=1}^{d} U_j \, ,
\end{equation}
its complexity is upper-bounded as
\begin{equation}
    \mathcal{C}(O) \leq \sqrt{d} \left( \sum_{j=1}^{d} \mathcal{C}(U_j) \right) \mathcal{C}_g \, .
\end{equation}

Lastly, we consider the more general linear combination of unitaries
\begin{equation}
\label{eq:linear_comb}
    O_{w} = \sum_{j=1}^{d} w_j U_j \,\,\, \text{with} \,\,\, \underset{i}{\sum}{w_i} = 1 \, .
\end{equation}
Returning to line (\ref{eq:READOFF}), recall that in
the standard LCU protocol \cite{Childs:2012gwh, Low:2016znh}, we read off the final state via:
\begin{equation}
\begin{split}
    \ket{\psi_F} &= \bra{0} \mathcal{P}_w^\dagger \left( \sum_{k=0}^{d-1} U_{k+1} \otimes \ket{k} \bra{k} \right) \mathcal{P}_w \ket{\psi_I} \otimes \ket{0} \\
    &= \sum_{j=1}^{d} w_{j} U_{j} \ket{\psi_I} \, .
\end{split}
\end{equation}
This gives an upper bound on the gate complexity for $O_{w}$
\begin{equation}
    \mathcal{C}\qty(O_{w}) \leq \frac{1}{\sqrt{p}} \left( \sum_{j=1}^{d} \mathcal{C}(U_j) \mathcal{C}_g + 2 \mathcal{C}\qty(\mathcal{P}_w) \right) \, ,
\end{equation}
where $p = \sum_i w_i^2$ is the probability for post-selection to succeed.\footnote{See footnote \ref{foot:DFWDFW}.}
Although we have included the contribution from the gate complexity of $\mathcal{P}_w$ here,
if the linear combination is over a small number of complex unitaries, this contribution can be neglected.

Our discussion above described one particular procedure to implement $\mathcal{X}$.
This is not necessarily the most efficient procedure. This is indicated by the fact that the procedure
only gives us an upper bound on the complexity of $\mathcal{X}$. In general, finding the optimal procedure
that computes the gate complexity is particularly difficult, even in the simple case of unitary operators.
Since we allow for a small error $\varepsilon$ in our implementation, one can easily conceive of other protocols
with lower gate complexity. For instance, the weights in \eqref{eq:linear_comb} can be approximated by fractions
with a common denominator
\begin{equation}
    \mathcal{X} = \sum_j \frac{n_j}{D} U_j + O(\varepsilon) \, .
\end{equation}
We can now use the simpler qudit protocol to implement $\mathcal{X}$.
The situation is further improved by the fact that the same unitary $U_j$
is associated with $n_j$ ancilla qudit states, so we can implement all of
them using a single sequence of controlled-$g$ gates.

\section{Distance Measure} \label{sec:DISTANCE}

In the previous section we argued that non-invertible symmetry operators can be interpreted as specifying
a class of quantum gate operations with LCUs. This specifies a computation with parallel quantum computers
which are then combined with a post-selection projection. Our aim in this section will be to quantify the proximity and complexity
of different quantum gate operations. We primarily focus again on the case of finite-dimensional Hilbert spaces,
though most of our considerations also extend to the infinite-dimensional setting once a suitable regularization is introduced.

Now, one of the awkward features of working with a fixed gate set is that it can be challenging to determine hard lower bounds on the gate complexity of a given
given operator. In the case of unitary gate operations, a
well-known way to address some of these issues is to to introduce a geometric measure on $U(2^N)$ for an $N$
qubit Hilbert space.\footnote{See in particular \cite{Nielsen:2005mkt, Nielsen:2006cea, Nielsen:2006mgv, Dowling:2006tnk, Gu:2008rsv},
and the overviews provided in \cite{Brown:2019rox, Acevedo:2025juf}.
The class of metrics proposed in these references have high curvature, but can be viewed as part of a broader family of
metrics / distance measures. For some examples of more general examples with lower curvature, see, e.g., reference
\cite{Brown:2022phc} where lower curvature choices are in the same \textbf{BQP} complexity class.}
See Appendix \ref{app:NIELSEN} for a brief review of Nielsen's complexity measure.
This formulation has been explored in the context of holographic systems in e.g., \cite{Brown:2015bva, Brown:2017jil}.

A priori, there are different notions of ``distance'' between operators one might entertain.
For example, given operators $A$ and $B$ of a $C^{\ast}$-algebra, the operator norm of the difference,
i.e., $\vert \vert A - B \vert \vert$ provides us with a notion of proximity between operators which is
distinct from the one defined by the Killing metric on a Lie group / symmetric space.

\begin{figure}[t!]
    \centering
    \includegraphics[trim={0 4cm 0 4cm}, scale = 0.5]{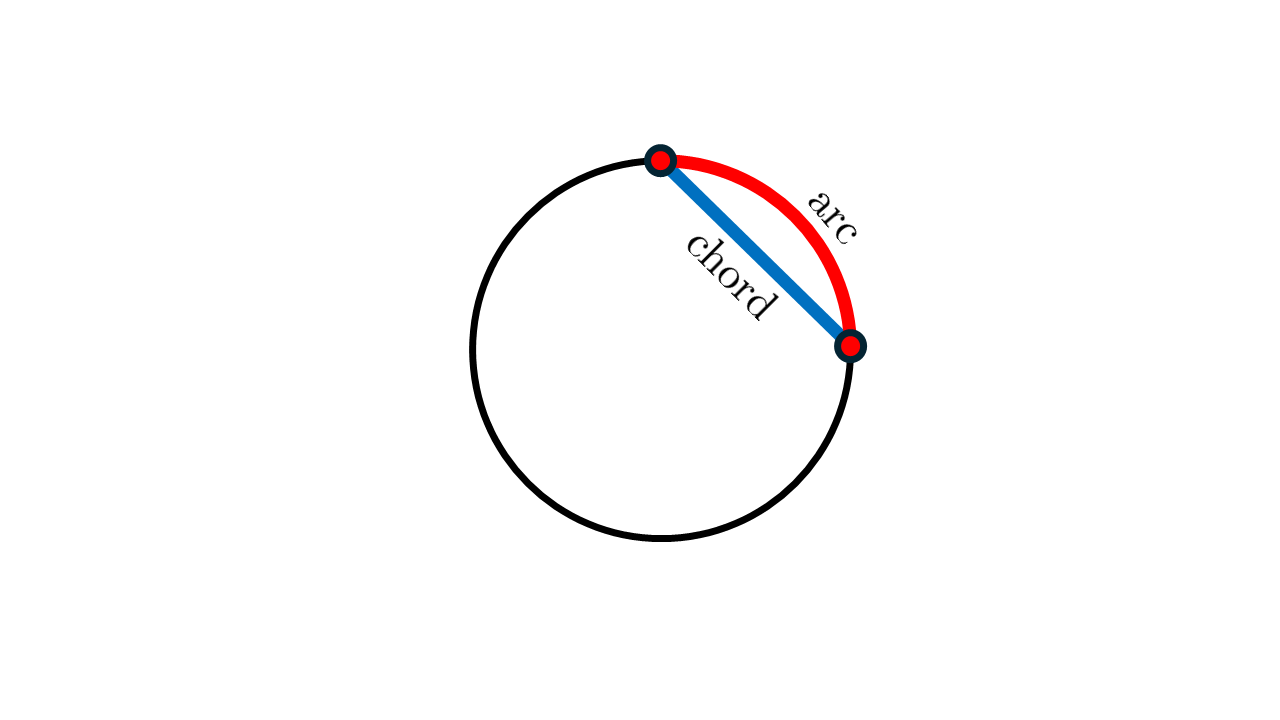}
    \caption{Depiction of the arc distance vs chord distance for operators, viewed as elements of $U(1)$,
    adapted from the figure in reference \cite{Brown:2022phc}. In the case of LCUs, these distinctions
    become blurred.}
    \label{fig:ChordVsArc}
\end{figure}

As an illustrative example \cite{Brown:2022phc}, consider different distance measures on $U(1)$.
One can either proceed along the arc of the circle, or instead embed the $S^1$ inside of $\mathbb{R}^2$ and construct the chord
that passes between these two points. One might view the arc length as more intrinsic to the group structure, but again this
depends on the computational question of interest. See figure \ref{fig:ChordVsArc}
(adapted from \cite{Brown:2022phc}) for a depiction of the difference between arc and chord lengths for $U(1)$.
This notion of arc versus chord length clearly generalizes to $U(2^N)$,
and different notions of distance provide different computational insights.
For a comparison of some natural choices, see e.g., \cite{Brown:2022phc}, and in particular table 2.

Let us now turn to the case of distance measures for LCUs.
One of the original motivations for this work was to come up with an appropriate distance
measure for non-invertible symmetry operators. Clearly, this requires extending beyond the standard group law. In particular, once
we start allowing linear combinations of unitary operators, the distinctions between ``arc length'' and ``chord length'' formulae become far less clear.
Faced with this, we shall instead adopt a more practical approach where we seek to define a natural notion of distance measure between
LCUs.

We now construct a distance measure which applies to both unitary gates as well as LCUs.
We focus on questions pertaining to distinguishability of a given gate and / or operator.
Our first aim will be to develop a uniform distance measure, analogous to working with a ``round'' group manifold. Working with the infinitesimal version of this distance measure, we then show how to produce a broader class of curved metrics which can be used to quantify different notions of gate complexities.

Our general goal will be to obtain a computationally tractable notion of
distance which admits an interpretation as specify a suitable complexity measure, akin to
what has been developed for purely unitary gate operations. Compared with the unitary case, we shall
already be presenting a generalization which introduces a reference mixed state $\rho$, which we
view as the ``typical'' state to stand in for a generic choice. In the purely quantum
information theoretic setting, one is often interested in optimization over some
particular choice of states to minimize, e.g., a distance. Our aim will be to
use such quantum information theoretic considerations as a motivating heuristic and interpretational
framework for understanding the structure of non-invertible symmetries in a quantum field theory. It is also worth noting that in practice, carrying out an explicit optimization can be quite unwieldy, and the aim here is to have in hand a computationally tractable way of measuring complexity.

\subsection{Round Distance Measures}

In this section we motivate a class of ``round'' distance measures, analogous to having a Killing metric. In practice, ``round'' will simply mean that all directions of motion in
the tangent space are equally favored. Later, we drop this consideration, proceeding to
generalizations suitable for broader gate complexity considerations.

Our starting point will rely on a suitable notion of proximity between density matrices, as specified by the trace distance. This a well-known measure of distance on the space of density matrices;
it is defined in terms of the trace norm, $||A||_1 = \mathrm{Tr} \sqrt{A^\dagger A}$. For density matrices $\rho$ and $\sigma$, the trace distance is given by
\begin{equation}
    D(\rho, \sigma) = \frac{1}{2} || \rho - \sigma ||_1  = \frac{1}{2} \mathrm{Tr} \sqrt{(\rho - \sigma)^2} \, .
\end{equation}
Note that $\sqrt{(\rho - \sigma)^2} \neq (\rho - \sigma)$ because $(\rho - \sigma)$ is not positive semi-definite i.e., it generically has both
positive and negative eigenvalues. We comment that while this is different from the relative entropy, in an infinitesimal limit, the two notions of proximity for full rank density matrices agree.\footnote{The relative entropy is defined as $S(\rho \vert \vert \sigma) = \mathrm{Tr} \rho \log \rho - \rho \log \sigma$. Expanding for infinitesimally different full rank density matrices results in the quantum Fisher information metric / Bures metric \cite{Bures:1969}. Yet another choice of distinguishability is the Fidelity $F(\rho \vert \vert \sigma ) = \left(\mathrm{Tr} \sqrt{\sqrt{\rho} \sigma \sqrt{\rho}} \right)^2$, which we can use to construct a modified distance $\widetilde{D}(\rho , \sigma) = \mathrm{arccos} \sqrt{F(\rho \vert \vert \sigma)}$. In the infinitesimal limit these different notions of proximity are all approximately equivalent.}
For pure states, the trace distance simplifies to
\begin{equation}
    D \left( \ket{\psi}\bra{\psi}, \ket{\phi}\bra{\phi} \right) = \sqrt{1 - | \langle \psi | \phi \rangle |^2} \, .
\end{equation}

We can now use the trace distance to devise a distance measure for arbitrary operators.
We comment that this not just the norm of the difference between operators, so in this sense
it retains the geometric ``flavor'' of an arc length formula used in complexity measure considerations.
For another distance measure based on the Frobenius norm, see Appendix \ref{app:FROBENIUS}.
These different choices all lead to qualitatively similar distances between operators.

Our distance will depend on an arbitrary normalized density matrix $\rho$.
Observe that under conjugation by an operator $\mathcal{X}$, the operator
$\mathcal{X} \rho \mathcal{X}^{\dag}$ may fail to be normalized, or may not be valid density matrix. To address these shortcomings, we shall instead work with the canonical purification of $\rho$ to a pure state $\ket{\psi_\rho}$ in the doubled Hilbert space $\mathcal{H} \otimes \mathcal{H}^*$.
Given an eigenbasis for $\rho$ i.e.,
\begin{equation}
    \rho = \sum_i p_i \ket{i}\bra{i} \, ,
\end{equation}
the canonically purified state is explicitly given by
\begin{equation}
    \ket{\psi_\rho} = \sum_i \sqrt{p_i} \ket{i}_L \otimes \ket{\overline{i}}_R \, .
\end{equation}
This purification is canonical in the sense that it does not depend on an explicit choice of basis.
Lastly, we can recover the original mixed state from the canonical purification by tracing out over one copy $\mathrm{Tr} _{\mathcal{H}^*} \qty(\ket{\psi_\rho}\bra{\psi_\rho}) = \rho$.

Starting from the canonically purified state $\ket{\psi_\rho}$, we obtain another pure state by acting with
$\mathcal{X}_L \otimes \mathbb{I}_R$. This pure state is not necessarily normalized; its norm squared is given by
\begin{equation}
    \bra{\psi_\rho} \left( \mathcal{X}_L \otimes \mathbb{I}_R \right)^\dagger \left( \mathcal{X}_L \otimes \mathbb{I}_R \right) \ket{\psi_\rho} = \sum_{i} p_i \bra{i} \mathcal{X}^\dagger \mathcal{X} \ket{i} = \mathrm{Tr}(\rho \mathcal{X}^\dagger \mathcal{X}) \, .
\end{equation}
Thus, the appropriately normalized state is
\begin{equation}
   \frac{\left( \mathcal{X}_L \otimes \mathbb{I}_R \right) \ket{\psi_\rho}}{\sqrt{\mathrm{Tr}(\rho \mathcal{X}^\dagger \mathcal{X})}} \, .
\end{equation}

We now introduce a distance between the identity operator and an operator $\mathcal{X}$ using the
trace distance for density matrices of the doubled Hilbert space $\mathcal{H}_{\mathrm{double}} = \mathcal{H} \otimes \mathcal{H}^{\ast}$:
\begin{equation}
\begin{split}
    \mathcal{D}_{\rho}(\mathbb{I}, \mathcal{X})
    &= D \left( \ket{\psi_\rho}\bra{\psi_\rho}, \frac{\left( \mathcal{X}_L \otimes \mathbb{I}_R \right) \ket{\psi_\rho} \bra{\psi_\rho} \left( \mathcal{X}_L \otimes \mathbb{I}_R \right)^\dagger}{\mathrm{Tr}(\rho \mathcal{X}^\dagger \mathcal{X})} \right) \\
    &= \sqrt{1 - \frac{\mathrm{Tr}(\rho \mathcal{X}^\dagger) \mathrm{Tr}(\rho \mathcal{X})}{\mathrm{Tr}(\rho \mathcal{X}^\dagger \mathcal{X})}} \, .
\end{split}
\end{equation}
More generally, we define the distance between two operators $\mathcal{X}$ and $\mathcal{Y}$ as
\begin{equation}
\label{eq:dist_def}
\begin{split}
    \mathcal{D}_{\rho}(\mathcal{X}, \mathcal{Y})
    &= D \left( \frac{\left( \mathcal{X}_L \otimes \mathbb{I}_R \right) \ket{\psi_\rho} \bra{\psi_\rho} \left( \mathcal{X}_L \otimes \mathbb{I}_R \right)^\dagger}{\mathrm{Tr}(\rho \mathcal{X}^\dagger \mathcal{X})} , \frac{\left( \mathcal{Y}_L \otimes \mathbb{I}_R \right) \ket{\psi_\rho} \bra{\psi_\rho} \left( \mathcal{Y}_L \otimes \mathbb{I}_R \right)^\dagger}{\mathrm{Tr}(\rho \mathcal{Y}^\dagger \mathcal{Y})} \right) \\
    &= \sqrt{1 - \frac{\mathrm{Tr}(\rho \mathcal{X}^\dagger \mathcal{Y}) \mathrm{Tr}(\rho \mathcal{Y}^\dagger \mathcal{X})}{\mathrm{Tr}(\rho \mathcal{X}^\dagger \mathcal{X}) \mathrm{Tr}(\rho \mathcal{Y}^\dagger \mathcal{Y})}} \, .
\end{split}
\end{equation}
In the special case of unitary operators, this simplifies to
\begin{equation}
    \mathcal{D}_{\rho}(U, V)
    = \sqrt{1 - \mathrm{Tr}(\rho U^\dagger V) \mathrm{Tr}(\rho V^\dagger U)} \, .
\end{equation}

As long as $\mathcal{X}$ and $\mathcal{Y}$ are not null operators,
$\mathrm{Tr} (\rho \mathcal{X}^\dagger \mathcal{X})$ and $\mathrm{Tr} (\rho \mathcal{Y}^\dagger \mathcal{Y})$
will be positive real values, because the product of $\rho$ and  $\mathcal{X}^\dagger \mathcal{X}$
is the product of two positive semi-definite matrices.
To streamline notation, we shall find it convenient to introduce an inner product notation, as induced by $\rho$.
Indeed, our trace induces a Hermitian inner product for operators so we opt to write:
\begin{equation}
\langle A, B \rangle \equiv \mathrm{Tr}(\rho A^{\dag} B),
\end{equation}
keeping the specific choice of $\rho$ implicit. In this notation, the distance reads as:
\begin{equation}
\begin{split}
    \mathcal{D}_{\rho}(\mathcal{X}, \mathcal{Y})
    = \sqrt{1 - \frac{\langle \mathcal{X} , \mathcal{Y} \rangle \langle \mathcal{Y} , \mathcal{X} \rangle}{\langle \mathcal{X},  \mathcal{X} \rangle \langle \mathcal{Y} , \mathcal{Y} \rangle}} \, .
\end{split}
\end{equation}

This distance inherits the usual properties of a distance measure from the trace distance.
Explicitly, we have
\begin{enumerate}
    \item $\mathcal{D}_\rho \qty(\mathcal{X}, \mathcal{X}) = 0$.

    \item $\mathcal{D}_\rho \qty(\mathcal{X}, \mathcal{Y}) \geq 0$.

    \item $\mathcal{D}_\rho \qty(\mathcal{X}, \mathcal{Y}) = \mathcal{D}_\rho \qty(\mathcal{Y}, \mathcal{X})$.

    \item $\mathcal{D}_\rho \qty(\mathcal{X}, \mathcal{Y}) \leq \mathcal{D}_\rho \qty(\mathcal{X}, \mathcal{O}) + \mathcal{D}_\rho \qty(\mathcal{O}, \mathcal{Y})$.
\end{enumerate}
Additionally, similar to the trace distance, it is also upper-bounded as $\mathcal{D}_\rho \qty(\mathcal{X}, \mathcal{Y}) \leq 1$. Moreover, the distance is invariant under scaling the operators by a constant i.e., $\mathcal{D}_\rho \qty(\lambda_1 \mathcal{X}, \lambda_2 \mathcal{Y}) = \mathcal{D}_\rho \qty(\mathcal{X}, \mathcal{Y})$ for any $\lambda_1, \lambda_2 \in \mathbb{C} \backslash \{0 \}$. As mentioned earlier, we treat operators that differ by an overall multiplicative constant to be equivalent, so this distance measure is consistent with the equivalence.

Note that for a distance measure, the inequality in the second condition must be strict when
$\mathcal{X} \neq \lambda \mathcal{Y}$ for any $\lambda \in \mathbb{C} \backslash \{0 \}$.
This is not guaranteed for
an arbitrary density matrix $\rho$, though if it is of full rank (e.g., the maximally mixed state), then the inequality is strict.\footnote{
As a simple example, where one can have equality even if $\mathcal{X}$ and $\mathcal{Y}$ are distinct, consider $\rho$ a pure state and any two operators
for which this pure state is an eigenstate.}

In addition to the usual properties of a distance measure, it is also left-invariant as needed
for any complexity measure \footnote{If one prefers a right-invariant measure instead, then one can change
$\mathcal{X}, \mathcal{Y} \to \mathcal{X}^\dagger, \mathcal{Y}^\dagger$ in the definition in \eqref{eq:dist_def}.}
\begin{equation}
\label{eq:left_inv}
    \mathcal{D}_{\rho}(\mathcal{O}_L \mathcal{X} , \mathcal{O}_L \mathcal{Y}) = \mathcal{D}_{\rho}(\mathcal{X} , \mathcal{Y}) \, ,
    \qquad\qquad
    \text{for any unitary } \mathcal{O}_L \, .
\end{equation}
Moreover, the distance is right-invariant if $\mathcal{O}_R$ commutes with $\rho$
\begin{equation}
\label{eq:right_inv}
    \mathcal{D}_{\rho}(\mathcal{X} \mathcal{O}_R, \mathcal{Y} \mathcal{O}_R) = \mathcal{D}_{\rho}(\mathcal{X} , \mathcal{Y}) \, ,
    \qquad\qquad
    \text{for any unitary } \mathcal{O}_R \text{ satisfying } [\rho, \mathcal{O}_R] = 0 \, .
\end{equation}
In particular, if we choose $\rho$ to be the maximally-mixed state,
i.e. $\rho \propto \mathbb{I}$, then the distance is bi-invariant.

One might ask whether one should bother choosing $\rho$ different from the maximally-mixed
state.\footnote{For infinite-dimensional systems one can formally view this as the infinite
temperature limit of the thermal density matrix $\rho_{\ther} = \exp(- \beta H)$.} We opt to include the choice of $\rho$ in our definition
because if one wants to distinguish between two operators by observing its action on a state randomly
drawn from a particular ensemble corresponding to $\rho$, then $\mathcal{D}_{\rho}(\bullet , \bullet)$ is a more suitable proxy for this distinguishability. In the particular case where the state is picked at random uniformly in the Hilbert space, $\rho$ should be the maximally-mixed state.
This is a good complexity measure, if one is completely agnostic to the state.
However, in most practical cases, the state being acted on is not completely random,
and this knowledge is incorporated into our definition by the inclusion of $\rho$ inside the trace.

\subsection{Infinitesimal Limit}

The definition we have given can be applied for any bounded operator which acts on the Hilbert space of states. In particular, we
can use it to provide a notion of proximity between symmetry operators, regardless of whether the symmetry
category in question has a continuum of simple objects, or is instead discrete.

As a cross-check on our definitions we now show that for operators expanded around a fixed reference operator $\mathcal{X}$, we obtain an infinitesimal distance which agrees, for example, with that of a continuous symmetry group. Consider,
then, the proximity between an operator $\mathcal{X}$ and an infinitesimal deformation $\mathcal{Y} = \mathcal{X} + d \mathcal{X}$.
We begin with the distance squared, i.e., the line element:
\begin{equation}
ds^2 = \mathcal{D}_{\mathcal{\rho}} (\mathcal{X} , \mathcal{Y})^2
= 1 - \frac{\langle \mathcal{X} , \mathcal{Y} \rangle \langle \mathcal{Y} , \mathcal{X} \rangle}{\langle \mathcal{X}, \mathcal{X} \rangle \langle \mathcal{Y} , \mathcal{Y} \rangle}.
\end{equation}
Expanding to quadratic order in the fluctuation $d \mathcal{X}$ yields:
\begin{equation}\label{eq:METRIC}
ds^2 = \frac{\langle d \mathcal{X} , d \mathcal{X} \rangle }{\langle \mathcal{X} , \mathcal{X} \rangle} - \left\vert \frac{\langle \mathcal{X} , d \mathcal{X} \rangle }{\langle \mathcal{X} , \mathcal{X} \rangle} \right\vert^{2}.
\end{equation}

As a special case, consider taking a maximally mixed state $\rho$ of a finite dimensional Hilbert space of dimension $K$ and
$\mathcal{X} = U$ a unitary operator. Then, $\langle \mathcal{X} , \mathcal{X}\rangle = 1$ and we obtain:
\begin{equation}\label{eq:METRICMETRIC}
ds^2 = \frac{1}{K}\mathrm{Tr} (d U^{\dag} d U) - \frac{1}{K^2} \mathrm{Tr}(U^{\dag} d U) \mathrm{Tr}(U d U^{\dag}),
\end{equation}
i.e., a metric on the projectivization $PU(K) = U(K) / U(1) $. Locally, this is the same as specifying the metric
for $SU(K)$.\footnote{Globally, we have that $U(K) = \frac{SU(K) \times U(1)}{\mathbb{Z}_K}$.}

\subsection{Generalized Complexity Metrics}

We now turn to a broader class of metrics which naturally extends the case of unitary
transformations to LCUs, with non-invertible symmetries as a special case.

To frame the discussion, let us first begin with some salient features of the unitary case. We first introduce a basis for the tangent space $T^{\ast}U(K)$, as spanned by $\sigma_{I}$.\footnote{In the case of an $N$ qubit system it is customary to set $K = 2^N$, but nothing requires us to specialize in this way.
When we do restrict in this way, a common choice is to take the basis $\{\sigma_I \}$ to be given by tensor products over the Pauli matrices and the identity, i.e., $\mathbb{I}_{2 \times 2}, \sigma_{x}, \sigma_{y}, \sigma_{z}$, but a priori any normalized spanning set will do.\label{foot:PAULI}}

Our primary requirement is that our local basis is orthonormal with respect to our inner product:
\begin{equation}
\langle \sigma_{I} , \sigma_{J} \rangle = \delta_{IJ}
\end{equation}
We raise and lower the local frame $I,J$ indices using $\delta_{IJ}$, for example
$\sigma_{I} = \delta_{IJ}\sigma^{J}$.

We can decompose the infinitesimal $U^{\dag} d U$ as:
\begin{equation}\label{eq:BASIS}
U^{\dag} d U = \mu^{I} \sigma_{I},
\end{equation}
where the $\mu^{I}$ are (infinitesimal) coefficients, and we have used an Einstein summation convention
for repeated indices. We can also extract the coefficient $\mu^{I}$ via:
\begin{equation}
\mu^{I} = \langle \sigma^{I} , U^{\dag} d U \rangle.
\end{equation}

Observe that for $U$ a unitary operator, our metric can also be written as:
\begin{equation}
\langle dU , d U \rangle = \langle U^{\dag} dU , U^{\dag} dU\rangle.
\end{equation}
Returning to our expansion in line (\ref{eq:BASIS}),
we now obtain:
\begin{equation}\label{eq:KILLING}
\langle dU , dU \rangle = \langle U^{\dag} dU , \sigma^{I} \rangle \delta_{IJ} \langle \sigma^{J}, U^{\dag} dU \rangle.
\end{equation}
We comment that normalization factors of ``$2^N$'' (see the discussion in Appendix \ref{app:NIELSEN})
are already accounted for by using a normalized density matrix.

The standard Killing metric corresponds to $\rho$ a maximally mixed state. The metrics of interest
for complexity considerations arise by dispensing with the explicit $\delta_{IJ}$ factor appearing
in line (\ref{eq:KILLING}) and substituting a more general tensor $\mathcal{I}_{IJ} = \delta_{IJ} \mathcal{I}(\sigma_{I})$:
\begin{equation}\label{eq:NIELSEN}
\langle dU^{\dag} , dU \rangle = \langle U^{\dag} dU , \sigma^{I} \rangle \mathcal{I}_{IJ} \langle \sigma^{J}, U^{\dag} dU \rangle.
\end{equation}
Here, $\mathcal{I}(\sigma_I)$ is a penalty factor associated with a particular direction. The
case of a round Killing metric amounts to no penalty at all,
i.e., $\mathcal{I}(\sigma_{I}) = 1$ for all $I$. In this setting, the distinction between
working with $U(K)$ versus its projectivization\footnote{See the discussion near line (\ref{eq:METRICMETRIC}).}
can be accounted for by a suitable choice of $\mathcal{I}_{IJ}$. As such, we
shall not dwell on this distinction further.

The discussion above generalizes the ``standard'' Nielsen complexity measure by incorporating a
choice of mixed state $\rho$ with respect to which we build an inner product $\langle \bullet , \bullet \rangle$. We now generalize this
further, extending this to the LCU / non-unitary / non-invertible symmetries case.
Indeed, by inspection of the infinitesimal ``round metric'' we found in line (\ref{eq:METRIC}), we see that the desired
generalization to the LCU setting is:
\begin{equation}
ds^2 = \frac{\langle \mathcal{X}^{\dag} d \mathcal{X} , \sigma^{I} \rangle \mathcal{I}_{IJ} \langle  \sigma^{J} , \mathcal{X}^{\dag} d \mathcal{X} \rangle}{\vert \langle \mathcal{X} , \mathcal{X} \rangle \vert^{2}},
\end{equation}
where the complexity factor $\mathcal{I}_{IJ}$ penalizes us the more we move off of a prescribed set of basis elements. We note that
in the case of finite / discrete non-invertible symmetries, we can simply view these operators as supplementing the collection of
unitary operators, i.e., we view them as a special choice of LCU. As such, we can always introduce an infinitesimal metric as above.

As a final comment, we note that while the round metric might appear to be ``too simple,'' it
lower bounds the complexity of a given operator; the more involved complexity metrics tend to further
stretch out the distance between operators. As such, finding a large distance with respect to the round Killing metric already indicates a high level of complexity.

\section{Distance for Non-Invertible Symmetries} \label{sec:EXAMPLES}

In the previous sections we showed that there is a natural distance measure for LCUs. This naturally generalizes
complexity measures used for unitary gates to a broader setting. In this section we specialize further and focus
on distance measures between non-invertible symmetry operators. This class of operators can be viewed in the quantum computational
setting as an additional set of gates which complement a standard gate set. That being said, our interest here will be more formal: we simply aim
to endow the space of non-invertible symmetries with a suitable notion of distance / proximity.

Now, in the case of non-invertible symmetries for quantum field theories, the underlying Hilbert space is infinite-dimensional. At a practical level,
we can opt to regularize this Hilbert space using a lattice discretization of the spatial component of the spacetime, e.g., as in \cite{Seiberg:2024gek}. This does introduce some lattice artifacts,
but provided we only ask about proximity up to some error / tolerance, this is a subleading issue.
The notion of distance / proximity we have introduced works equally well regardless of whether we are dealing with a continuous family of non-invertible symmetries or a discretized collection.
In the former case, we have a clear generalization of the metric for a Lie group or a symmetric space. In the case of a discretized collection, we recognize that we can form linear combinations of operators, and so ``fill in'' the discretized symmetry operators by more general LCUs. In all cases, then, we can speak of a manifold with topology induced from our distance measure / metric.

\begin{figure}[t!]
    \centering
    \includegraphics[width=0.75\textwidth]{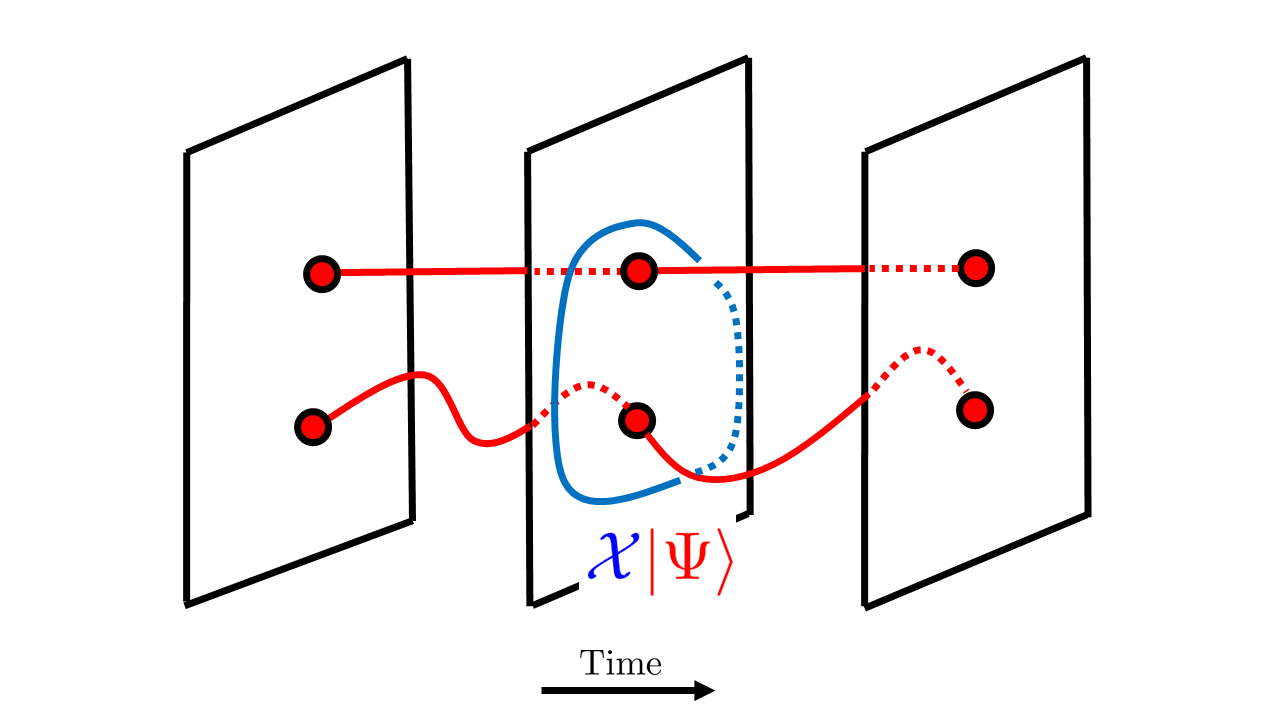}
    \caption{Depiction of the $(D+1)$-dimensional bulk Symmetry Theory / Symmetry Topological Field Theory Hilbert space
    interpretation of symmetry operators for a $D$-dimensional QFT. A bulk topological operator $\mathcal{X}$ acts on a state $\vert \Psi \rangle$
    of the bulk. In this depiction the radial direction serves as a time coordinate for a Euclidean boundary system.}
    \label{fig:SymTFT}
\end{figure}

\begin{figure}[t!]
    \centering
    \includegraphics[width=0.75\textwidth]{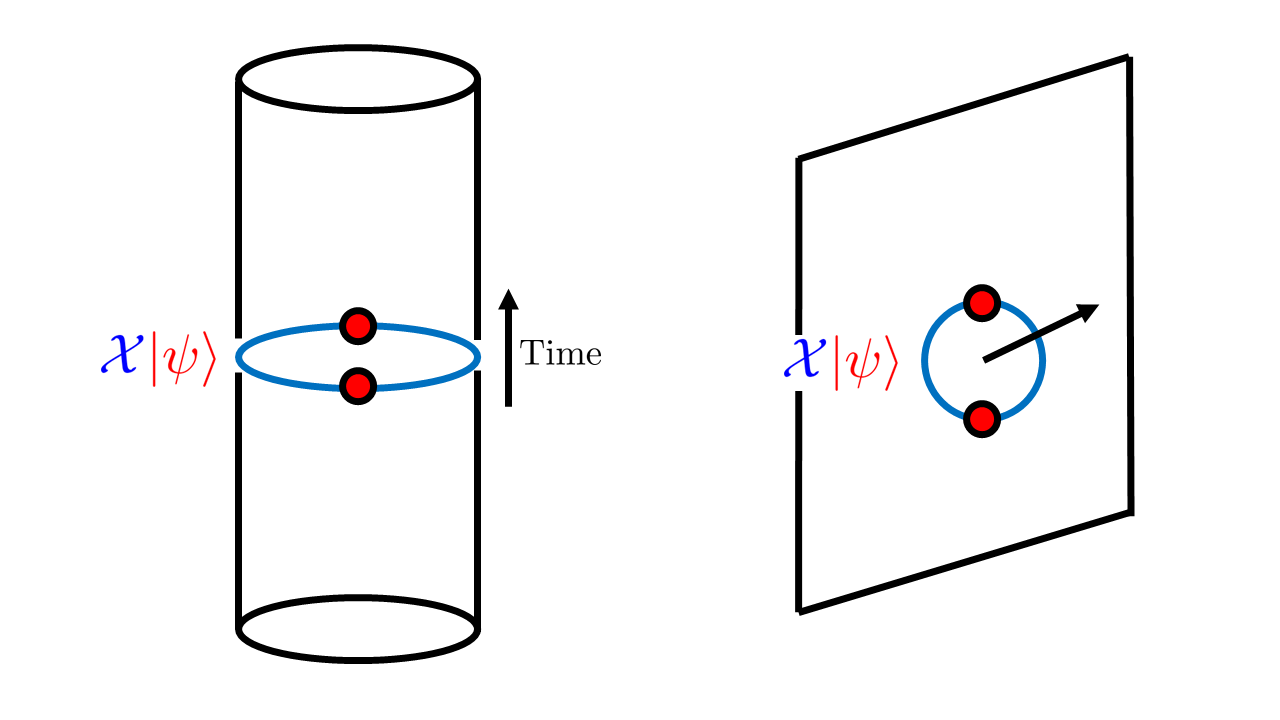}
    \caption{Depiction of topological symmetry operator $\mathcal{X}$ acting on a state $\vert \psi \rangle$ of a $D$-dimensional
    absolute QFT, quantized on the spacetime $\mathbb{R} \times S^{D-1}$ (left). For a conformal field theory, this is equivalent to
    working with a theory defined on $\mathbb{R}^{D}$ in which we radially quantize the theory (right).}
    \label{fig:CylinderAction}
\end{figure}

It is also worth noting
that there is actually another Hilbert space interpretation available, given by a Symmetry Theory / Symmetry TFT for the symmetries of a
QFT.\footnote{See for example \cite{Reshetikhin:1991tc,Turaev:1992hq, Barrett:1993ab, Witten:1998wy, Fuchs:2002cm, Kirillov2010TuraevViroIA,
Kapustin:2010if, Kitaev:2011dxc, Fuchs:2012dt, Freed:2012bs, Kong:2014qka, Kong:2017hcw, Heckman:2017uxe, Freed:2018cec,
Gaiotto:2020iye, Kong:2020cie, Apruzzi:2021nmk, Freed:2022qnc, Kaidi:2022cpf, Antinucci:2022vyk, Lawrie:2023tdz, Baume:2023kkf, Yu:2023nyn,
Brennan:2024fgj,Antinucci:2024zjp, Argurio:2024oym, Franco:2024mxa, Heckman:2024zdo, Gagliano:2024off, Cordova:2024iti, Cvetic:2024dzu,
Bhardwaj:2024igy, Bonetti:2024cjk, Apruzzi:2024htg, Yu:2024jtk, Jia:2025jmn, Apruzzi:2025mdl, Heckman:2025lmw, Luo:2025phx, Torres:2025jcb,
Bergman:2026lnz} and references therein for a partial list of references to foundational early work, as well as more recent generalizations.}
In this setting, one decompresses the absolute QFT to a higher-dimensional system in which there is a
relative QFT on one end, and a choice of boundary conditions on the other end enforcing the global form of the symmetries in the absolute QFT.
In this bulk theory we can perform canonical quantization with a timelike direction radially running from one boundary to the other.
Then, the symmetry operators in question simply act on the Hilbert space of states obtained with possible anyons / defects
inserted. For finite symmetry categories, the resulting anyon Hilbert spaces (i.e., for a fixed number of defects)
are finite-dimensional, and there is a natural interpretation in terms of operators
acting on states (and thus serving as quantum gates) in this bulk theory. See figure \ref{fig:SymTFT} for a depiction of the $(D+1)$-dimensional bulk Symmetry Theory / Symmetry TFT
Hilbert space interpretation, and figure \ref{fig:CylinderAction}
for a depiction of the $D$-dimensional QFT interpretation.

That being said, we shall primarily focus on the interpretation of our non-invertible symmetries directly in the absolute QFT$_D$. Our aim will be to
study some illustrative examples; we defer a more exhaustive treatment to future work.

Our first example will be 4D $O(2)$ Maxwell theory, an example of a $D > 2$ non-invertible
symmetry treated in detail in \cite{Heidenreich:2021xpr}.
After this, we turn to non-invertible symmetries in certain 2D CFTs, first considering some examples
for RCFTs, and then for the symmetric orbifold of a 2D CFT. In all cases, we calculate a
distance between symmetry operators, illustrating how our notion of proximity naturally extends the one present for Lie groups and symmetric spaces. An interesting result of our analysis is that
the ``simple objects'' of a symmetry category can be \textit{computationally} complex.

\subsection{Example: $O(2)$ Gauge Theory}

Our first example is 4D $O(2)$ gauge theory. Recall that in Maxwell theory, the gauge group is
$U(1)$ and there are continuous electric and magnetic 1-form symmetry operators \cite{Gaiotto:2014kfa}. For our purposes, we can present these as:\footnote{Some care is needed in properly defining the operator $\mathcal{U}_{\alpha}[\Sigma]$ in terms of the integral over the dual field strength $\widetilde{F}$, since in an electric operator basis the Hodge star operation will appear. The more precise statement is that we construct the operator by ensuring that it acts appropriately on the magnetic line operators. That said, such subtleties will play little role in the discussion to follow.}
\begin{equation}
\mathcal{U}_{\alpha}[\Sigma] =  e^{i \alpha \widetilde{F}_{\Sigma}}\,\,\, \text{and} \,\,\,
\mathcal{V}_{\beta}[\Sigma] =  e^{i \beta F_{\Sigma}},
\end{equation}
for $\alpha , \beta \in [0,2\pi)$ (we absorb all physical couplings into our conventions for the field strength).
Here, to ease the notation we have introduced the integral of the field strengths over
$\Sigma$ as $F_{\Sigma}$ and $\widetilde{F}_{\Sigma}$.
These symmetry operators are manifestly topological, and obey a group-like fusion rule. Suppose, however, that we now gauge the
charge conjugation symmetry of Maxwell theory. Then, the relevant gauge group is $O(2) = \mathbb{Z}_{2}^{C} \ltimes U(1)$.
There is a linear combination which is invariant under charge conjugation, and this is the operator
which survives in the new theory:\footnote{It is worth noting that although we have written the
$\mathcal{X}$'s as sums of ``unitary operators,'' these unitaries
are not actually operators of the ``quotient theory.'' If we supplement the
$C$-invariant Hilbert space by a twisted sector with $C$-odd states, then such an interpretation would be appropriate, but
there is nothing inconsistent with remaining with just the $C$-invariant states.
Nevertheless, since we have a bounded operator acting on our Hilbert space, there
does exist a formal decomposition of our operators into a sum of unitary operators,
as per our discussion in section \ref{sec:GATES}.}
\begin{equation}
\mathcal{X}_{\alpha} = \mathcal{U}_{\alpha} + \mathcal{U}_{\alpha}^{\dag} \,\,\, \text{and} \,\,\, \mathcal{Y}_{\beta} = \mathcal{V}_{\beta} + \mathcal{V}_{\beta}^{\dag},
\end{equation}
where we keep the supporting surface $\Sigma$ implicit. Observe that in the $O(2)$ gauge theory $\alpha \in [0 , \pi]$ and $\beta \in [0 , \pi]$.
By inspection, the fusion rules for these operators have more than one summand,
so we obtain an example of a non-invertible symmetry in $D > 2$.

Let us now compute the distance between our operators, focusing for ease of exposition on the electric 1-form symmetry.
Recall that our general expression for the metric is:
\begin{equation}
ds^2 = \frac{\langle d \mathcal{X} , d \mathcal{X} \rangle}{\langle \mathcal{X} , \mathcal{X} \rangle} - \left \vert \frac{\langle \mathcal{X} , d \mathcal{X} \rangle}{\langle \mathcal{X} , \mathcal{X} \rangle} \right \vert^{2}.
\end{equation}
It therefore suffices to compute the various inner products which appear in our general formula. To this end,
we first compute the inner product:
\begin{equation}
\langle \mathcal{X}_{\alpha}, \mathcal{X}_{\alpha}\rangle  = \mathrm{Tr}(\rho \mathcal{X}^{\dag}_{\alpha} \mathcal{X}_{\alpha}) = 2 + \langle \mathbb{I} , \mathcal{X}_{2 \alpha} \rangle.
\end{equation}
Next, observe that the local fluctuation in the parameter $\alpha$ yields:
\begin{equation}
d \mathcal{X}_{\alpha} = \left(i \widetilde{F}_{\Sigma} e^{i \alpha \widetilde{F}_{\Sigma}} - i \widetilde{F}_{\Sigma} e^{- i \alpha \widetilde{F}_{\Sigma}}\right) d \alpha,
\end{equation}
expressed in terms of operators in the parent $U(1)$ gauge theory. As such, we extract the inner product:
\begin{equation}
\langle d \mathcal{X}_{\alpha} , d \mathcal{X}_{\alpha} \rangle =  \langle \widetilde{F}_{\Sigma}^{2}, 2 \mathbb{I} - \mathcal{X}_{2 \alpha} \rangle d \alpha^2.
\end{equation}
Lastly, we need the inner product between $\mathcal{X}_{\alpha}$ and $d \mathcal{X}_{\alpha}$:
\begin{equation}
\langle \mathcal{X}_{\alpha} , d \mathcal{X}_{\alpha} \rangle = \langle \widetilde{F}_{\Sigma} ,
i e^{2 i \alpha \widetilde{F}_{\Sigma}} - i e^{- 2i \alpha \widetilde{F}_{\Sigma}} \rangle d \alpha.
\end{equation}
Using the form of the metric presented in line (\ref{eq:METRIC}), we obtain:
\begin{equation}
ds^2 = e^{2 h(\alpha)} d\alpha^2,\,\,\,\text{with}\,\,\, e^{2h(\alpha)} = \frac{\langle \widetilde{F}_{\Sigma}^{2}, 2\mathbb{I} - \mathcal{X}_{2 \alpha} \rangle}{ 2 + \langle \mathbb{I} , \mathcal{X}_{2 \alpha} \rangle} -
\left\vert \frac{\langle \widetilde{F}_{\Sigma} ,
e^{2 i \alpha \widetilde{F}_{\Sigma}} - e^{- 2i \alpha \widetilde{F}_{\Sigma}} \rangle}{2 + \langle \mathbb{I} , \mathcal{X}_{2 \alpha} \rangle} \right\vert^2
\end{equation}
All told, then, we get the (unsurprising) result that the metric involves a function of $\alpha$, as well as the choice of
density matrix $\rho$. The function $e^{2h(\alpha)}$ is in general non-trivial (i.e., non-zero).\footnote{For a holographic example involving
thermal expectation values of a symmetry operator, see \cite{Heckman:2025isn}.}
By inspection, we also see that $e^{2 h(\alpha)}$ vanishes at $\alpha = 0$.

\subsection{Example: RCFTs}

Perhaps one of the best studied examples of non-invertible symmetries are the Verlinde lines \cite{Verlinde:1988sn} appearing in 2D rational conformal field theories (RCFTs). The main simplification in dealing with an RCFT is that we have a finite number of primary operators, and thus a corresponding finite list of Verlinde lines. See e.g., references \cite{Zuber:2000ia, Cappelli:2009xj, Chang:2018iay} for additional details on RCFTs, and \cite{Frohlich:2004ef, Frohlich:2006ch, Feiguin:2006ydp, Chang:2018iay, Thorngren:2019iar} for recent discussion on the structure of non-invertible symmetries in such 2D CFTs.\footnote{The literature here is vast so we only offer an incomplete set of references. See \cite{Shao:2023gho} and references therein for a more complete treatment.}

On a torus with modular parameter $\tau$, the partition function for an RCFT is given by the finite sum:
\begin{equation}
    Z = \underset{j,\overline{j}}{\sum} N_{j \bar{j} } \chi_j (q) \chi_{\bar{j}} (\bar{q}), \qquad q = e^{2 \pi i \tau}, \ \bar{q} = e^{- 2 \pi i \bar{\tau}},
\end{equation}
where $\chi_j$ are the characters for the primaries. To avoid unnecessary complications in what follows we therefore focus on diagonal RCFTs, i.e., where $N_{j \bar{j}} = \delta_{j,\overline{j}}$.

In the diagonal models, the Verlinde lines \cite{Verlinde:1988sn} are in one to one correspondence with the primary operators. The action of a Verlinde line on a primary state is:\footnote{In non-diagonal models the action of the Verlinde lines can mix states in different sectors according to the fusion algebra \cite{Petkova:2000ip}, and the correspondence with primary operators is also more involved.  See e.g., \cite{Thorngren:2019iar} for a recent discussion}
\begin{equation}
    \mathcal{L}_n \ket{i} = \frac{S_{ni}}{S_{0i}} \ket{i},
\end{equation}
where $S$ is the modular S-matrix of the 2D CFT.\footnote{For a choice of modular parameter $\tau$, the S-matrix
elements are implicitly defined by $\chi_{i}(-1/\tau) = \underset{j}{\sum} S_{ij} \chi_{j}(\tau)$, in the obvious notation.}

The fusion rule for the Verlinde lines is:\footnote{Compared with the notation in previous sections,
we now adopt the standard conventions in the 2D literature, i.e., since we are dealing with a line operator we write $\mathcal{L}$ to denote our non-invertible symmetry operator.}
\begin{equation}
\mathcal{L}_{i} \mathcal{L}_{j} = \underset{k}{\sum} N_{ij}^{k} \mathcal{L}_{k},
\end{equation}
where the fusion coefficients obey the Verlinde relation \cite{Verlinde:1988sn}:
\begin{equation}
N_{ij}^{k} = \underset{\ell}{\sum} \frac{S_{i\ell} S_{j \ell} S^{\ast}_{k \ell}}{S_{1 \ell}},
\end{equation}
where $S_{ij}$ denotes the entries of the modular S-matrix.

We shall be interested in using our distance measure to calculate the proximity between a pair of Verlinde lines.
We focus on the distance measure as specified by working with the
finite temperature thermal state:
\begin{equation}
    \rho_{\ther} = \sum_i e^{-\beta H} \ket{\psi_i}\bra{\psi_i},
\end{equation}
where the Hamiltonian is $H = \frac{2 \pi}{L} \qty(L_0 + \bar{L}_0 - c/12)$, and we
identify $\tau = i\beta/L$ (i.e., no twist / angular momentum).
The trace distance between two Verlinde lines $\mathcal{L}_n$ and $\mathcal{L}_m$ is
\begin{equation}
    \mathcal{D}_{\ther} \qty(\mathcal{L}_n , \mathcal{L}_m) = \sqrt{1 - \frac{\mathrm{Tr} \qty(\rho_{\ther} \mathcal{L}^\dagger_n \mathcal{L}_m)
    \mathrm{Tr} \qty(\rho_{\ther} \mathcal{L}^\dagger_m \mathcal{L}_n)}{
    \mathrm{Tr} \qty(\rho_{\ther} \mathcal{L}^\dagger_n \mathcal{L}_n)
    \mathrm{Tr} \qty(\rho_{\ther} \mathcal{L}^\dagger_m \mathcal{L}_m)
    }}.
\end{equation}
The trace $\mathrm{Tr} \qty(\rho_{\ther} \mathcal{L}^\dagger_n \mathcal{L}_m) $ evaluates to:
\begin{equation}
    \mathrm{Tr} \qty(\rho_{\ther} \mathcal{L}^\dagger_n \mathcal{L}_m) = \sum_i \qty(\frac{S_{ni}}{S_{1i}})^* \qty(\frac{S_{mi}}{S_{1i}}) \bra{\psi_i} e^{-\beta H} \ket{\psi_i}.
\end{equation}
This can be reorganized into sums over Verma modules resulting in:
\begin{equation}
    \mathrm{Tr} \qty(\rho_{\ther} \mathcal{L}^\dagger_n \mathcal{L}_m) = \sum_a \qty(\frac{S_{na}}{S_{1a}})^* \qty(\frac{S_{ma}}{S_{1a}}) \abs{\chi_a \qty(i\beta/L)}^2 .
\end{equation}
Our plan will be to evaluate the trace distance to leading order in both the low and high temperature limits.

This is essentially as far as we can go without specializing further. To this end, we now pick a few representative examples.
As examples, we treat the case of the $\widehat{\mathfrak{su}}(2)_k$ WZW model, as well as the unitary $(A_{p}, A_{p^{\prime}})$ minimal models.

\subsubsection{$\widehat{\mathfrak{su}}(2)_k$ WZW Model}

As a first class of examples, we can consider the $\widehat{\mathfrak{su}}(2)_k$ WZW Model.
We begin by stating our conventions. Recall that this model has central charge $c = 3k / k+2$, with primary operators
$\vert j \rangle$ for $j = 1,...,k + 1$ of (left-moving) conformal weight $h_j = (j - 1)(j+1)/4(k+2)$.
The modular S-matrix has entries: \begin{equation}
S_{nm}= \sqrt{\frac{2}{k+2}} \sin \frac{nm \pi}{k+2}
\end{equation}
for $1, \leq n,m \leq k+1$. The primary $\vert j \rangle$ is in correspondence with a representation of $SU(2)$ of dimension $j$. Indeed,
in the $k \rightarrow \infty$  limit, the target space becomes that of a geometric $S^{3}$.

We shall be interested in computing the trace distance at both high and low temperatures. At high temperature, we essentially have the maximally mixed
state, while at low temperature, (i.e., $\beta \rightarrow \infty$ and $q \rightarrow 0$) the density matrix
will be dominated by the lowest weight primaries $\chi_1$ and $\chi_2$.

Consider first the low temperature limit. The terms in each sum over Verma modules will be dominated by the lowest weight states:
\begin{equation}
\begin{split}
    \mathrm{Tr} \qty( \rho_{\ther} \mathcal{L}^{\dag}_n \mathcal{L}_m)
    &\approx \qty(\frac{S_{n1}}{S_{11}})^* \qty(\frac{S_{m1}}{S_{11}}) \abs{\chi_1 (i\beta/L)}^2 +
    \qty(\frac{S_{n2}}{S_{12}})^* \qty(\frac{S_{m2}}{S_{12}}) \abs{\chi_2 (i\beta/L)}^2 \\
    &= \frac{S_{n1}^* S_{m1}}{\abs{S_{11}}^2}\abs{\chi_1 (i\beta/L)}^2
    \qty(1 + \frac{S_{n2}^* S_{m2} \abs{S_{11}}^2} {S_{n1}^* S_{m1} \abs{S_{12}}^2}  \abs{\frac{\chi_2 (i\beta/L)}{\chi_1(i\beta/L)}}^2 ).
\end{split}
\end{equation}
We now define the quantity
\begin{equation}
    \varepsilon \equiv \abs{\frac{S_{11} \chi_2 (i\beta/L)}{S_{12} \chi_1 (i\beta/L)}} \approx \frac{1}{\cos \frac{\pi}{k+2}} e^{- \frac{3\pi}{2(k+2) L} \beta }  .
\end{equation}
Expanding the distance in terms of $\varepsilon \ll 1$, we find that
\begin{equation}
    \mathcal{D}(\mathcal{L}_{n}, \mathcal{L}_m)  = \varepsilon \abs{ \frac{S_{m2}}{S_{m1}} - \frac{S_{n2}}{S_{n1}} }.
\end{equation}
Inserting the expression for the S-matrix elements, we obtain the final result:
\begin{equation}
    \mathcal{D}(\mathcal{L}_{n}, \mathcal{L}_m) = 2 \varepsilon \abs{\cos \frac{m \pi}{k+2} - \cos \frac{n \pi}{k+2}}.
\end{equation}

Consider next the high temperature limit $\beta \rightarrow 0$. In this limit, we can use the modular S-transformation to express the character as:
\begin{equation}
    \chi_i (\tau) = \sum_j S_{ij} \chi_j(-1/\tau).
\end{equation}
The sum becomes dominated by the lowest weight term at high temperature:
\begin{equation}
    \chi_i (\tau) \approx S_{i1} \chi_1(-1/\tau).
\end{equation}
Inserting this back into the trace, the character can be pulled out of the sum, and it cancels between the numerator and denominator. Using $S^\dagger S = \mathbb{I}$, it follows that
\begin{equation}
    \mathcal{D} \qty( \mathcal{L}_n, \mathcal{L}_m) =
    1 - \delta_{nm}.
\end{equation}
So, the distance is either 0 if $n = m$ or maximally far apart if $ n \neq m$.

\begin{figure}[t!]
    \centering
    \includegraphics[width=0.75\textwidth]{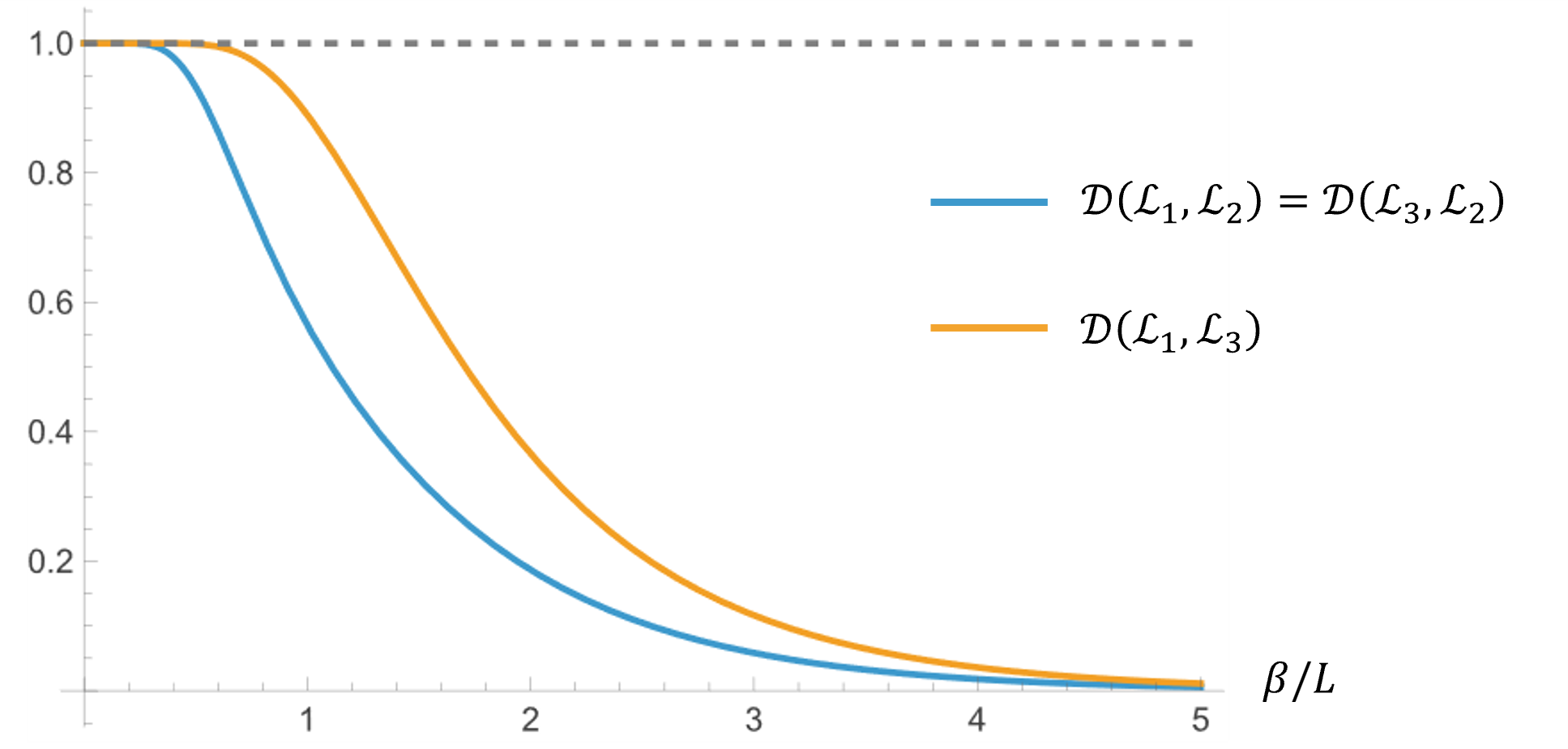}
    \caption{Distance between the $\mathcal{L}_{m}$ operators in the $\widehat{\mathfrak{su}}(2)_{k}$ CFT at level $k = 3$.}
    \label{fig:A3_CFT}
\end{figure}

\begin{figure}[t!]
    \centering
    \includegraphics[width=0.75\textwidth]{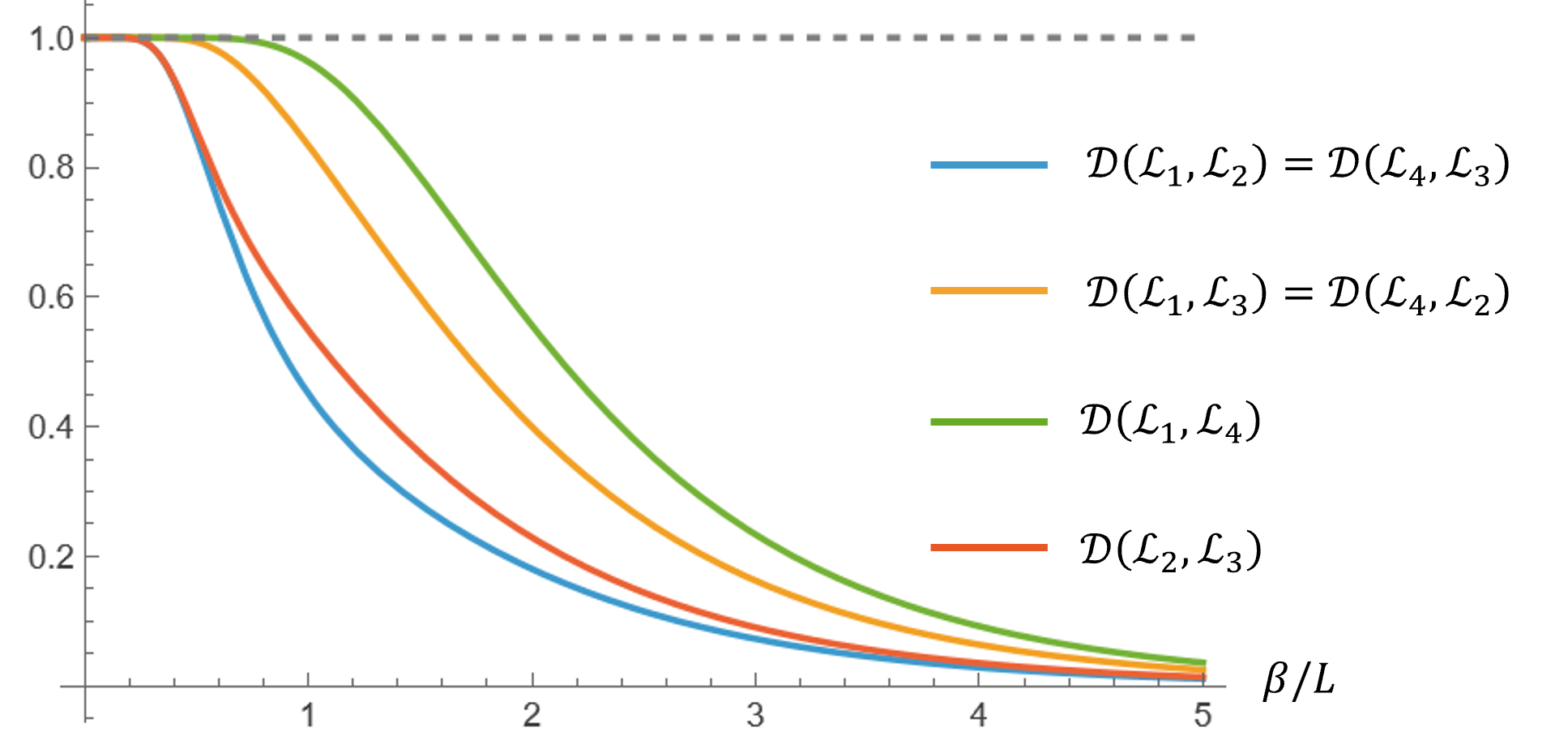}
    \caption{Distance between the $\mathcal{L}_{m}$ operators in the $\widehat{\mathfrak{su}}(2)_{k}$ CFT at level $k = 4$.}
    \label{fig:A4_CFT}
\end{figure}

In figures \ref{fig:A3_CFT} and \ref{fig:A4_CFT}
we respectively show some examples of distance measures at levels $k = 3 $ and $k = 4$.

\subsubsection{The $(A_p, A_{p^{\prime}})$ Minimal Models}

As another example, we consider the $(A_p, A_{p'})$ minimal model, also known as the $(p,p')$ minimal model.
Here $p$ and $p'$ are co-prime with $p>p'$.
The central charge of this theory is:
\begin{equation}
    c = 1 - \frac{6(p-p')^2}{p p'}.
\end{equation}
The primary states are labeled as $\vert r , s \rangle $ with $1 \leq r \leq p - 1$ and $1 \leq s \leq p' - 1$.
In our conventions, the conformal weight is:
\begin{equation}
h_{rs} = \frac{(ps - p'r)^2 - (p - p')^2}{4 p p'},
\end{equation}
The characters are given by
\begin{equation}
    \chi_{(r,s)}(\tau) = \frac{1}{\eta(\tau)} \sum_{j \in \mathbb{Z}} \left[ q^{(2pp'j+pr-p's)^2/4pp'} - q^{(2pp'j+pr+p's)^2/4pp'} \right]
\end{equation}
where the $(r,s)$-indices satisfy the restrictions $1 \leq r \leq p'$, $1 \leq s \leq p$, and $p's < pr$.
The S-matrix elements are given by
\begin{equation}
    S_{(r_1,s_1),(r_2,s_2)} = 2 \sqrt{\frac{2}{pp'}} (-1)^{1 + r_1 s_2 + r_2 s_1} \sin \frac{\pi p r_1 r_2}{p'} \sin \frac{\pi p' s_1 s_2}{p} .
\end{equation}
In what follows, we restrict to the unitary case where $p' = p-1$. Observe that in this case, the weights simplify:
\begin{equation}
h_{rs} = \frac{(p(s - r) + r)^2 - 1}{4 (p^2 - p)},
\end{equation}
and the lowest weight states are $(r,s) = (1,1)$ and $(r,s) = (2,2)$.

As earlier, in the low temperature limit, the distance is dominated by the lowest dimension primaries -- the $(1,1)$ and $(2,2)$ primaries. It follows that:
\begin{equation}
    \mathcal{D} \qty(\mathcal{L}_{(r_1,s_1)}, \mathcal{L}_{(r_2,s_2)}) \approx \varepsilon \abs{\frac{S_{(r_1,s_1),(2,2)}}{S_{(r_1,s_1),(1,1)}} - \frac{S_{(r_2,s_2),(2,2)}}{S_{(r_2,s_2),(1,1)}}} ,
\end{equation}
where
\begin{equation}
    \varepsilon = \abs{\frac{S_{(1,1),(1,1)} \chi_{(2,2)}(i\beta/L)}{S_{(1,1),(2,2)} \chi_{(1,1)}(i\beta/L)}} \approx \frac{1}{4 \cos \frac{\pi}{p-1} \cos \frac{\pi}{p}} e^{ - \frac{3\pi}{2(p^2-p) L} \beta}.
\end{equation}
Using the explicit S-matrix elements, we obtain the final result in the low temperature limit:
\begin{equation}
    \mathcal{D}\qty(\mathcal{L}_{(r_1,s_1)}, \mathcal{L}_{(r_2,s_2)}) \approx 4 \varepsilon \abs{\cos \frac{\pi r_1}{p-1} \cos\frac{\pi s_1}{p} - \cos \frac{\pi r_2}{p-1} \cos\frac{\pi s_2}{p}} .
\end{equation}

In the high temperature limit, the results are identical to the previous case -- the distance is either 0 if the operators match or 1 if the operators are distinct.

\begin{figure}[t!]
    \centering
    \includegraphics[width=0.8\textwidth]{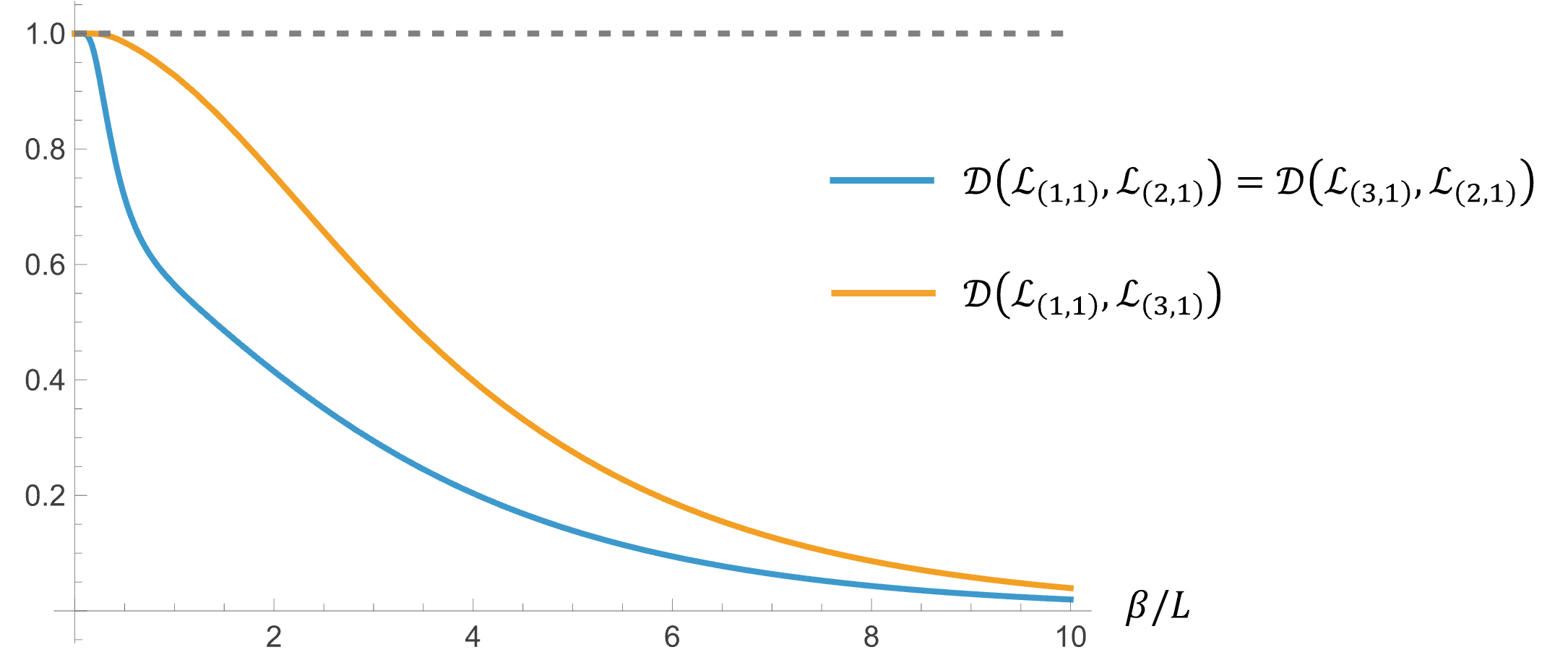}
    \caption{Distance between the $\mathcal{L}_{(r,s)}$ operators in the Ising CFT, $(p,p')=(4,3)$.}
    \label{fig:Ising_CFT}
\end{figure}

\subsection{Example: Symmetric Orbifold CFT}

As a final example, we consider the $N$-fold product of a 2D CFT. Given a seed theory $\mathcal{C}$,
we construct the $N$-fold symmetric product $\mathrm{Sym}^{N} \mathcal{C} = \mathcal{C}^N / S_N$, where
$S_N$, is the symmetric group on $N$ letters. This
construction plays a prominent role in the D1-D5 system \cite{Strominger:1996sh, Seiberg:1999xz}, as well as the tensionless limit of the AdS$_3$ / CFT$_2$ correspondence \cite{Gaberdiel:2018rqv, Eberhardt:2018ouy}. Denote by $Z(\tau, \overline{\tau})$ the partition function for the seed theory, and that of the $N$-fold symmetric product by $Z_N(\tau , \overline{\tau})$.
They are related via the grand canonical generating function
\begin{equation}
    \sum_{N} p^N Z_N(\tau, \overline{\tau}) = \exp \left[ \sum_{m \in \mathbb{Z}^+} p^m T_m Z(\tau, \overline{\tau}) \right] \, ,
\end{equation}
where $T_k$ is the Hecke operator that acts as
\begin{equation}
    T_m Z(\tau) = \frac{1}{m} \sum_{d|m} \sum_{j=1}^{d-1} Z \qty(\frac{m \tau + jd}{d^2}, \frac{m \overline{\tau} + jd}{d^2}) \, .
\end{equation}

This theory enjoys a $\mathrm{Rep} (S_N)$ global symmetry, so we have symmetry operators naturally labeled by representations of $S_N$, i.e., we write $\mathcal{L}_{i}$ for $i \in \mathrm{Rep}(S_N)$. The fusion algebra for these symmetry operators is obtained by taking the tensor product of representations of $S_N$.

The primary operators are likewise naturally labeled by conjugacy classes of $S_{N}$, and these are in turn specified by a choice of partition of $N$, i.e., they are labeled by Young diagrams. Our aim will be to calculate the distance between $\mathcal{L}_{i}$ and
$\mathcal{L}_{j}$ in both the low and high temperature limit, i.e., we set $\tau = i \beta / L$, in the obvious notation.

Consider first the low temperature limit, i.e., $\beta \rightarrow \infty$.
It suffices to consider just the conjugacy classes for the lowest weight sectors, i.e., the identity, $[1]$, as well as the
2-cycle $[2]$, i.e., transpositions. All other contributions are suppressed in the low temperature limit. Setting $\tau = i \beta / L$, we have:
Using $Z(\tau , \overline{\tau}) \approx e^{\frac{\pi c}{6 L} \beta}$ with $c$ the central charge of the seed theory, we get:
\begin{equation}
    \mathrm{Tr}( \rho_{\ther} \mathcal{L}_i \mathcal{L}_j) \approx \chi_{i} ([e])\chi_j([e]) \left( e^{\frac{\pi c}{6 L} N \beta} + \dots \right) + \chi_i([2]) \chi_j([2]) \left( e^{\frac{\pi c}{6 L} \left( N - \frac{3}{2} \right) \beta} + \dots \right) \, ,
\end{equation}
and hence
\begin{equation}
\label{eq:dist_orbifold_low_temp}
    \mathcal{D}( \mathcal{L}_i , \mathcal{L}_j ) \approx \abs{\frac{\chi_i([2])} {\chi_i([e])} - \frac{\chi_j([2])} {\chi_j([e])}} e^{-\frac{\pi c}{8 L} \beta} \, .
\end{equation}

Next, we consider the high temperature limit  $\beta \to 0$.
Modular invariance of $Z(\tau, \overline{\tau})$ implies that $Z(\tau , \overline{\tau}) \approx e^{\frac{\pi c L}{6} \beta^{-1}}$.
At some values of $N$, it is possible that the ratio of characters in \eqref{eq:dist_orbifold_low_temp} are the same for particular values of $i,j$ (e.g., $N=6$). In that case, we include the contribution from the next lowest sector corresponding to $[3]$.
It follows that
\begin{equation}
    \mathrm{Tr}( \rho_{\ther} \mathcal{L}_i \mathcal{L}_j) \approx e^{\frac{\pi c L}{6} N \beta^{-1}} \delta_{ij} \, ,
\end{equation}
where we have used the Schur orthogonality of the irreducible characters.
Evidently,
\begin{equation}
    \mathcal{D}( \mathcal{L}_i , \mathcal{L}_j ) \approx 1 - \delta_{ij} \, .
\end{equation}
As earlier, the distance is either 0 if the operators match or 1 if the operators are distinct.

\begin{figure}[t!]
    \centering
    \includegraphics[width=0.8\textwidth]{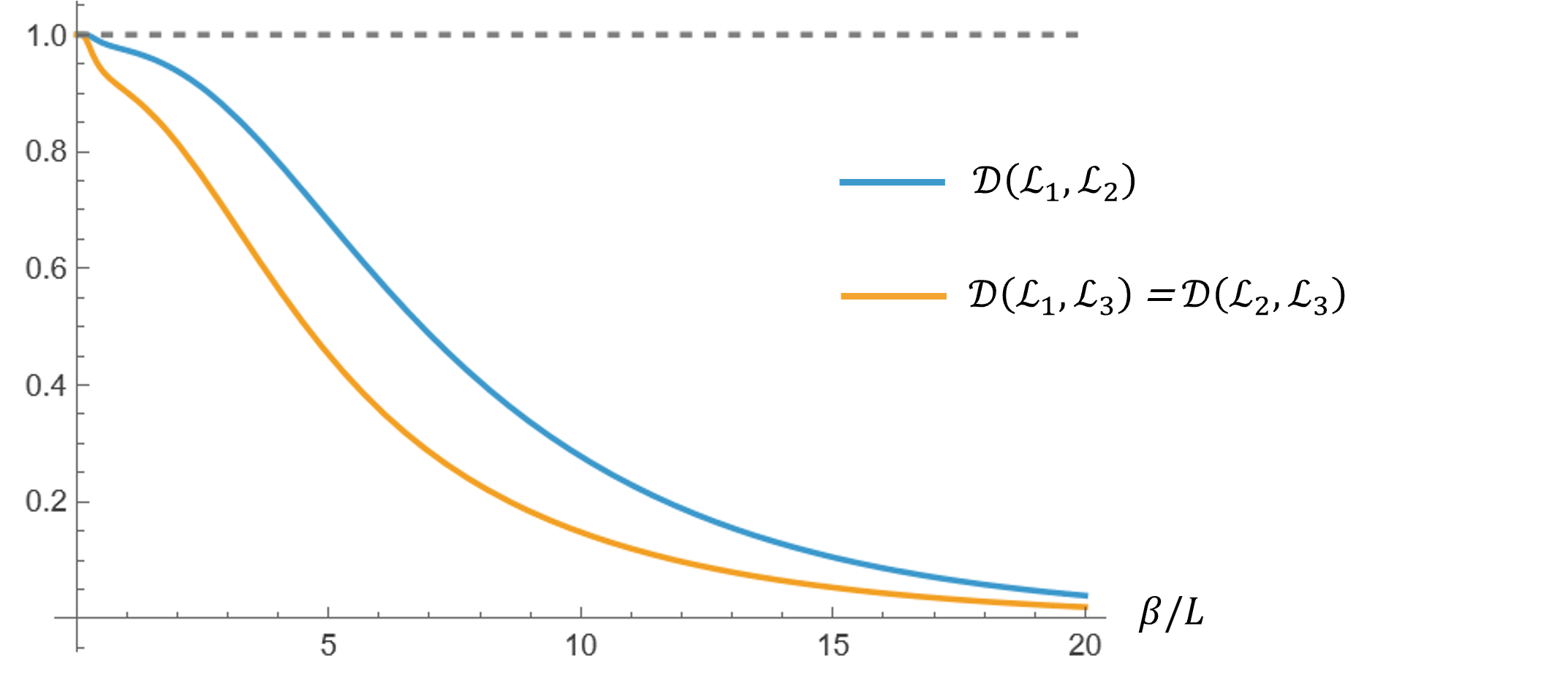}
    \caption{Distance between the $\mathcal{L}_{i}$ operators in the $S_3$ orbifold CFT with the seed theory being a free fermion CFT.
    Here, the line operator $\mathcal{L}_{1}$ corresponds to the trivial representation, $\mathcal{L}_{2}$ to the signed representation,
    and $\mathcal{L}_3$ corresponds to the standard representation.}
    \label{fig:S3_CFT}
\end{figure}

\subsection{Simple Objects can be Maximally Complex}

Observe that in the high temperature limit for $\rho_{\mathrm{th}}$, which corresponds to the maximally mixed state, the distance between symmetry operators is as large as possible; we recover a
discrete topology. Since the round Killing metric tends to serve as a lower bound on the complexity of an operator, we conclude that non-invertible symmetries, i.e.,
the ``simple objects'' of a symmetry category are \textit{computationally} rather complex, i.e., for a generic gate
set, they cannot easily be simulated in terms of a small number of gates.

\section{Conclusions} \label{sec:CONC}

In this paper, we have shown that non-invertible symmetries can be interpreted as a special class of LCUs, that is, linear combinations of unitary operators. In a truncated approximation of the Hilbert space we can visualize their action on states as specifying parallelized quantum computations which are then combined via post-selection.
There is a corresponding gate complexity for such quantum gates, and we have shown there is also a natural complexity / accuracy distance
measure which extends the case of distances on unitaries for an $N$ qubit system to general LCUs. We have also computed the distance / proximity
between non-invertible symmetries in some examples where these structures feature prominently. In the remainder of this section,
we discuss some future directions of investigation.

In much of this work we made the mild assumption that a suitable regulator / error tolerance could
be introduced in order to obtain a suitable approximation for various gate sets. Especially
in the context of a quantum field theory where the Hilbert space dimension is
infinite-dimensional, it would be desirable to dispense with such regulators altogether.
Along these lines, it would be interesting to see whether the recent results of \cite{Sorce:2025usc}
could be used to construct a suitable distance measure.

One of the original motivations for this work was to better understand the role of non-invertible symmetries
in holographic systems, i.e., by taking operators defined in a SymTFT sliver and pushing them into the gravitational bulk,
as in \cite{Heckman:2024oot, Heckman:2025lmw}.
From this perspective, one natural use for the distance measure introduced here would be to interrogate a broader notion of
bulk complexity, especially in systems with a bulge in the bulk (e.g., a Python's lunch configuration as in \cite{Brown:2019rox}).
We have seen that the simple objects of our symmetry category are often computationally very complex. It would be interesting
to study this property further, especially in holographic systems, where the onset of computational complexity leads to rather striking
phenomena. See e.g., \cite{Brown:2015bva, Carmi:2016wjl, Balasubramanian:2019wgd} for various holographic treatments of complexity.
Indeed, the appearance of a linear combination of unitary operators suggests a picture in which parallel computations / universes proceed but are then
recombined by a post-selection observer at some prescribed time. It would be interesting to study further examples along these lines.

The appearance of a weighted sum over unitary operators is also interesting in the context of constructing concrete realizations of disorder averaged
quantities, i.e., by taking multiple copies of a system and then summing them together. A top down realization of one such construction is given in
\cite{Heckman:2021vzx}, and it would be interesting to revisit this construction in light of the present work. More broadly, averaging over symmetry operators has been considered in a number of different contexts (see e.g., \cite{Cordova:2022rer, Cummings:2025zfe}) and it would be interesting to revisit these results in light of the present work.

For spontaneous breaking of group-like symmetries, the metric on a coset space plays an important
role in the construction of the effective action for the Goldstone modes.\footnote{For a recent discussion of Goldstone bosons associated with spontaneous breaking of a continuous non-invertible symmetry, see reference \cite{GarciaEtxebarria:2022jky}.} We anticipate that likewise, having a well-posed distance for non-invertible symmetries will play a similar role in the construction of the corresponding effective field theories.


\section*{Acknowledgements}

We thank C. Cummings, S.N. Meynet, J. Sorce, and X. Yu for helpful comments on an earlier draft.
We also thank V. Chakrabhavi, C. Cummings, S.N. Meynet, J. Sorce, and X. Yu for helpful discussions.
JJH thanks the Kavli IPMU for hospitality during part of this work.
JJH thanks the organizers and administrative staff for the KITP meeting
``Generalized Symmetries in Quantum Field Theory: High Energy Physics, Condensed Matter, and Quantum Gravity'' for rejecting his application.
JJH thanks the 2025 Simons Physics Summer Workshop for hospitality during part of this work.
JJH thanks the Simons Collaboration on Global Categorical Symmetries
Annual Meeting 2025 for hospitality during part of this work.
The work of JJH and CM  is supported by DOE (HEP) Award DE-SC0013528. The work of JJH is also supported by
BSF grant 2022100 and a University Research Foundation grant at the University of Pennsylvania.
The work of RJH is supported by an NSF Graduate Research Fellowship under Grant No. DGE-2236662.

\newpage

\appendix

\section{Nielsen Complexity} \label{app:NIELSEN}
Nielsen complexity, or geometric complexity, is a way of defining the complexity of a quantum operation using the geometric notion of distance on a manifold. It quantifies the cost of implementing a unitary operation in a continuous way as opposed to counting the number of discrete gates needed to generate a transformation by identifying the operation with a geodesic with respect to an appropriate metric.

As developed in \cite{Nielsen:2005mkt}, for some set of exactly universal $N$-qubit unitary gates given by $\mathcal{G}$, there is an exponential mapping of to a set of Hermitian matrices $\mathcal{H}$. The group generated by $\mathcal{G}$ should be $U(2^N)$. One can then denote the pairs of operators and tangent vectors, $(V,H)$ where $V \in U(2^N)$ and $H$ defines a tangent vector in $T_V U(2^N)$. A ``local metric" $F$ can then be used to quantify the difficulty of implementing $H$ at the point $V$. The local metric must satisfy the following properties:

\begin{itemize}
    \item Continuity
    \item Positivity
    \item Positive homogeneity
    \item Achievement of the infimum
    \item The triangle inequality
    \item Right-invariance
\end{itemize}

This local metric can be used to define a Finsler metric on $U(2^N)$, which is a generalization of a Riemannian metric. Suppose now we want to identify a path between the identity $\mathbb{I}$ and a
unitary operator $U$. The function defining this path $\gamma (t)$ must satisfy
\begin{equation}
    \frac{\\d \gamma}{\\d t} = -i H(t) \gamma, \quad \gamma(0) = \mathbb{I}, \quad \gamma(1) = U,
\end{equation}
where $t$ is on some interval $I = [0,1]$.
Given a distance measure $F$, the complexity of $U$ is then the infimum of the length of this path. It can be precisely defined by the norm of the tangent vectors along the path from $\mathbb{I}$ to $U$:
\begin{equation}
    \mathcal{D}_{F} \lparen U\rparen
    \coloneqq \mathcal{D}_{F} \lparen \mathbb{I}, U \rparen = \inf_\gamma \int^1_0 ||\gamma^\prime (s)||_F \ \dd s .
\end{equation}
We say that the distance measure $F$ is ``$\mathcal{G}$-bounding'' when
$F(V,H) \le 1$. This distance provides a lower bound on the gate complexity.

In the context of Lie groups the simplest metric to define is the Killing metric. Elements of this metric can then be scaled up or down to describe the complexity of operations in a system. We work with respect to a prescribed basis of the tangent space $\sigma_{I}$.\footnote{See footnote \ref{foot:PAULI}.}
Following the discussion in \cite{Brown:2022phc}, the Killing metric for $U(2^N)$ in the generalized Pauli basis can be expressed as:
\begin{equation}
    ds^2 = \sum_{I,J} (i \overline{\Tr} \sigma_I U dU^\dagger)\delta_{IJ}(i \overline{\Tr} \sigma_J U dU^\dagger),
\end{equation}
where we have introduced the normalized trace $\overline{\Tr} = 2^{-N} \Tr$.

This form does not have any penalty associated with interactions of more than two qubits. By replacing $\delta_{IJ}$ with a diagonal penalty matrix $\mathcal{I}_{IJ} = \delta_{IJ} \mathcal{I} (\sigma_I)$, which generalizes the Killing metric to a Finsler metric, this information a can be described. Moving in directions associated with interactions involving more than two qubits is more difficult.
We comment that if the distance specified by the symmetric Killing metric is already large, then the distance defined by $F$ is typically even bigger, i.e., a large Killing metric distance already
indicates high complexity.

\section{Frobenius Norm Distance} \label{app:FROBENIUS}

The Frobenius norm of a matrix $A$ is defined as
\begin{equation}
    || A ||_{\mathfrak{F}} = \sqrt{\mathrm{Tr}(A^\dagger A)} = \sqrt{\sum_{i,j=1}^{n} |A_{ij}|^2} \, .
\end{equation}
The Frobenius norm induces a distance measure on the space of matrices as:
\begin{equation}
    D_{\mathfrak{F}} (A,B) = || A - B ||_{\mathfrak{F}}
\end{equation}
This distance satisfies the usual properties of a distance measure.
Explicitly, we have
\begin{enumerate}
    \item $D_{\mathfrak{F}} \qty(A, A) = 0$.

    \item $D_{\mathfrak{F}} (A, B) \geq 0$ if $U \neq V$.

    \item $D_{\mathfrak{F}} (A, B) = D_{\mathfrak{F}} (B, A)$.

    \item $D_{\mathfrak{F}} (A,B) \leq D_{\mathfrak{F}} \qty(A, C) + D_{\mathfrak{F}} \qty(C, B)$.
\end{enumerate}

We can use this Frobenius norm to define an alternative distance measure for unitary operators
\begin{equation}
\label{eq:frob_dist_unitary}
    \mathcal{D}_{\mathfrak{F},\rho} \qty(U,V) = D_{\mathfrak{F}} \qty(U \sqrt{\rho}, V \sqrt{\rho}) = || (U - V) \sqrt{\rho} ||_{\mathfrak{F}} \, .
\end{equation}
where $\rho$ is an arbitrary density matrix.
Note that the inclusion of this matrix in our definition of distance ensures that our distance is properly normalized as
\begin{equation}
    ||U \sqrt{\rho}||_{\mathfrak{F}} = \mathrm{Tr}( \rho U^\dagger U) = 1 \, .
\end{equation}
Using triangle inequality, it follows that this distance is upper bounded as
\begin{equation}
    \mathcal{D}_{\mathfrak{F},\rho} (U,V) \leq ||U \sqrt{\rho}||_{\mathfrak{F}} + ||V \sqrt{\rho}||_{\mathfrak{F}} =  2 \, .
\end{equation}
This bound is tight as $V = -U$ saturates it.
This distance measure is left-invariant and is also right-invariant if $W_R$ commutes with $\rho$ i.e., it satisfies \eqref{eq:left_inv} and \eqref{eq:right_inv}.

If we choose $\rho$ to be the maximally mixed state, this distance becomes the normalized Frobenius norm
\begin{equation}
    \mathcal{D}_{\mathfrak{F},\rho} \qty(U,V) = || U-V ||_{\overline{{\mathfrak{F}}}} = \sqrt{\overline{\mathrm{Tr}} \big( (U-V)^\dagger (U-V) \big)} \, ,
\end{equation}
where $\overline{\mathrm{Tr}}(\cdot)$ is the normalized trace that satisfies $\overline{\mathrm{Tr}}(\mathbb{I}) = 1$.
Clearly, this distance is bi-invariant.

We can extend this distance to more general operators as follows
\begin{equation}
    \mathcal{D}_{\mathfrak{F},\rho} \qty(U,V) = D_{\mathfrak{F}} \qty( \frac{1}{\sqrt{\mathrm{Tr}(\rho U^\dagger U)}} \sqrt{\rho} U ,  \frac{1}{\sqrt{\mathrm{Tr}(\rho V^\dagger V)}} \sqrt{\rho} V ) \, .
\end{equation}
Since $\mathrm{Tr}(\rho U^\dagger U) = 1$ when $U$ is unitary, this distance reduces to \eqref{eq:frob_dist_unitary} for unitaries.
Lastly, we can express this distance entirely in terms of traces as
\begin{equation}
    \mathcal{D}_{\mathfrak{F},\rho} \qty(U,V) = \sqrt{2 -  \frac{\mathrm{Tr}(\rho U^\dagger V) + \mathrm{Tr}(\rho V^\dagger U)}{\sqrt{\mathrm{Tr}(\rho U^\dagger U) \mathrm{Tr}(\rho V^\dagger V)}}} \, .
\end{equation}
The distances we obtain using this new distance measure are qualitatively similar to the trace distance.
For the RCFT and symmetric orbifold CFT examples, the low temperature values exactly match the values we obtained in section \ref{sec:EXAMPLES}.
In the high temperature limit, the distance is either $0$ if the operators match or $\sqrt{2}$ if the operators are distinct.

\newpage

\bibliographystyle{utphys}
\bibliography{NonInvDist}

\end{document}